# From truth to computability II

Giorgi Japaridze[*]

## Abstract

Computability logic is a formal theory of computational tasks and resources. Formulas in it represent interactive computational problems, and "truth" is understood as algorithmic solvability. Interactive computational problems, in turn, are defined as a certain sort games between a machine and its environment, with logical operators standing for operations on such games. Within the ambitious program of finding axiomatizations for incrementally rich fragments of this semantically introduced logic, the earlier article "From truth to computability I" proved soundness and completeness for system **CL3**, whose language has the so called parallel connectives (including negation), choice connectives, choice quantifiers, and blind quantifiers. The present paper extends that result to the significantly more expressive system **CL4** with the same collection of logical operators. What makes **CL4** expressive is the presence of two sorts of atoms in its language: *elementary atoms*, representing elementary computational problems (i.e. predicates, i.e. problems of zero degree of interactivity), and *general atoms*, representing arbitrary computational problems. **CL4** conservatively extends **CL3**, with the latter being nothing but the general-atom-free fragment of the former. Removing the blind (classical) group of quantifiers from the language of **CL4** is shown to yield a decidable logic despite the fact that the latter is still first-order.

## Contents

[*]This material is based upon work supported by the National Science Foundation under Grant No. 0208816

# 1 Introduction

Being a continuation of [6], this article fully relies on the terminology, notation, conventions and technical results of its predecessor, with which the reader is assumed to be familiar.

The atoms of our old friend **CL3** represent predicates rather than computational problems in general. So, **CL3** only allows us to talk about elementary problems and their particular ($\neg, \land, \lor, \rightarrow, \sqcap, \sqcup, \sqcap, \sqcup, \forall, \exists$)-combinations. This is a rather serious limitation of expressive power. By far not every natural problem of potential interest can be expressed as a combination of the above sort, including all problems with infinitely or arbitrarily long legal runs. And even though finite strict games such as chess always can be, in principle, represented as $\sqcap, \sqcup$-combinations of elementary games (terminal positions), the sizes of such representations would often be beyond reasonable. To get a feel of what is missing in **CL3**, imagine a situation where we just want to be able to directly say "*Chess*", "*Go*" etc. (i.e. have these as atoms), and then ask questions such as whether and how the problem (of winning) $Go \sqcup (Chess \land Checkers)$ can be reduced to $(Go \sqcup Chess) \land (Go \sqcup Checkers)$. **CL3** does not give us this ability because its formalism does not allow atoms for non-elementary problems. True, **CL3** still might help us find a positive answer to at least the 'whether' part of the above particular question. Specifically, in view of the soundness of **CL3**, its would suffice to show that, for *all* formulas $E, F, G$, **CL3** proves $(E \sqcup F) \land (E \sqcup G) \rightarrow E \sqcup (F \land G)$. This, however, would require some creative reasoning in the metatheory of **CL3** rather than **CL3** itself. And even if we managed to succeed in this metareasoning, we still would not be able to automatically conclude that every problem of the form $A \sqcup (B \land C)$ is algorithmically reducible to $(A \sqcup B) \land (A \sqcup C)$. Rather, we would only know that this is so as long as $A, B, C$ are ($\neg, \land, \lor, \rightarrow, \sqcap, \sqcup, \sqcap, \sqcup, \forall, \exists$)-combinations of elementary problems. While the class of problems of this sort is certainly interesting and nontrivial, it — as already pointed out — is only a modest fraction of the collection of all entities that we call interactive computational problems.

By simply redefining the semantics of the language of **CL3** and no longer requiring that its atoms be interpreted as elementary problems, we would certainly gain a lot. But perhaps just as much would be lost: the class of valid formulas would shrink, victimizing many important and innocent principles. The point is that elementary problems are meaningful and interesting in their own right, and losing the ability to differentiate them from problems in general would be too much of a sacrifice. What else would be automatically gone is the nice fact that classical logic is a fragment of the new logic.

Computability logic has a better solution. It simply allows two sorts of atoms in its language, one for elementary problems and the other for all problems. This way, not only do we have the ability to characterize valid principles for problems of either sort within the same formal system, but we can as well capture principles that intermix elementary problems with not-necessarily-elementary ones. Such an approach also has technical advantages. As we are going to see, logic **CL4**, whose language extends that of **CL3** by adding to it the second sort of atoms, has a rather simple axiomatization, while it remains unclear whether there is a reasonable deductive system for the fragment of computability logic whose language only has the second sort of atoms.[1]

This article is primarily devoted to a soundness and completeness proof for the above-mentioned system **CL4**. Its secondary result is a proof of the decidability of the $\forall, \exists$-free (yet first-order) fragment of **CL4**. These results, with a forward reference to the present paper, have been announced in [3, 8]. Soundness and completeness for the propositional fragment **CL2** of **CL4** was proven in [5]. The present article strengthens the results of its predecessor in the same sense as [5] strengthens [4], the difference being that here we deal with the dramatically more expressive first-order level, as opposed to the propositional level of [4, 5].

Mathematical curiosity aside, the main practical import of our results is first of all related to the potential of basing applied theories or knowledge base systems on **CL4**, the latter being a reasonable, computationally meaningful alternative to classical logic. The non-logical axioms — or knowledgebase — of a **CL4**-based applied system/theory would be any collection of (formulas representing) problems whose algorithmic solutions are known. Then our soundness result for **CL4**, which comes in a strong form called uniform-constructive soundness, guarantees that every theorem $T$ of the theory also has an algorithmic solution and that, furthermore, such a solution, itself, can be effectively constructed from a proof of $T$. This makes **CL4** a systematic problem-solving tool: finding a solution for a given problem reduces to finding a proof of that problem in a **CL4**-based theory, with completeness meaning that, in its language, **CL4** is as perfect/strong as as a formal tool could possibly be. Section 6 of [6] discussed some of the advantages and appeal of basing applied systems on **CL3**.[2] That discussion, however, dealt with systems where the new logical operators (those of computability logic) were applied to the old non-logical atoms (those of classical logic): predicates. **CL4** offers a more far-reaching way for enriching traditional applied systems, allowing us to include in their vocabularies atoms that represent higher-level entities: computational problems of arbitrary complexities and degrees of interactivity, including infinite dialogues between a problem-solving agent and its environment. This is a substantially new level of expressiveness that the needs of advanced interactive applied systems would be inherently calling for.

---

[1]The very recently conceived syntactic approach, called *cirquent calculus* [7], does provide a nice axiomatization for a certain fragment of computability logic in a language that only has the second sort of atoms. The language of that fragment is only a modest fraction of that of **CL4** though, and, while the cirquent calculus approach seems promising, at this point **CL4** remains by far the most expressive system for computability logic known to be sound and complete.

## 2 Logic CL4

Let us rename what in [6] was called "predicate letters" into "**elementary letters**". The language of **CL4** augments that of **CL3** with an additional sort of syntactic objects called **general letters**, again each such object coming with a fixed arity. "Letter" is a common name for elementary and general letters. It is assumed that, for every $n \geq 0$, there are infinitely many $n$-ary letters of either sort. All general letters are considered non-logical. An **atom** of the language of **CL4** is $L(t_1, \ldots, t_n)$, where $L$ is an $n$-ary letter and each $t_i$ is a *term*, i.e. (as before) an element of the set of *variables* $\{v_0, v_1, \ldots\}$ or *constants* $\{0, 1, \ldots\}$. This atom is said to be $n$-**ary** because $L$ is so. In a similar way, we extend the usage of the terms "elementary", "general", "non-logical" etc. from letters to atoms. **CL4-formulas**, or henceforth often simply **formulas**, are built from atoms in the same way as **CL3**-formulas, using the "propositional" connectives $\neg, \wedge, \vee, \rightarrow, \sqcap, \sqcup$ and quantifiers $\forall, \exists, \sqcap, \sqcup$. The definition of a *free occurrence* of a variable $x$ is as before: this means that the occurrence is not in the scope of $\forall x, \exists x, \sqcap x$ or $\sqcup x$. When a given occurrence of a variable $x$ is in the scope of $\mathcal{Q}x$ for more than one quantifier $\mathcal{Q}$, the occurrence is considered bound by the quantifier "nearest" to it. E.g., the occurrence of $x$ in $\forall x(R \sqcup \sqcup x P(x))$ is bound by $\sqcup$ rather than $\forall$. An occurrence bound by $\forall$ or $\exists$ is said to be **blindly bound**.

An **interpretation** for the language of **CL4** is a function that sends each $n$-ary general (resp. elementary non-logical) letter $L$ to a problem (resp. elementary problem) with an attached tuple $(x_1, \ldots, x_n)$ of $n$ pairwise distinct variables. We denote such a problem by $L^*(x_1, \ldots, x_n)$, call $(x_1, \ldots, x_n)$ the **canonical tuple** of $L^*$, and say that "* **interprets** $L$ as $L^*(x_1, \ldots, x_n)$". There is no redundancy in saying so, for this phrase conveys nontrivial information about the context-setting canonical tuple — a context in which, according to our earlier conventions (Section 4 of [6]), we can unambiguously write $L^*(t_1, \ldots, t_n)$ to mean $L^*[x_1/t_1, \ldots, x_n/t_n]$. Sometimes, however, there is no need in being specific about the canonical tuple. In such cases we may simply write "$L^*$", which should be strictly understood as an abbreviation for $L^*(x_1, \ldots, x_n)$ where $(x_1, \ldots, x_n)$ is the canonical tuple of $L^*$.

Note that, just as in the case of **CL3**, we do not insist that interpretations respect the arities of letters. Specifically, we do not require that $L^*$ depend on only (or all) the variables from its canonical tuple, or even be finitary. To avoid the possibility of unpleasant collisions of variables under such a liberal approach, in [6] interpretations were restricted to "admissible" ones. Some additional caution is necessary in the present case, as we need to guarantee that $\forall x$ and $\exists x$ are only applied to games for which they are defined, i.e. $x$-unistructural games. Therefore the concept of admissibility now sharpens as follows:

**Definition 2.1** For a formula $F$ and interpretation $^*$, we say that $^*$ is $F$-**admissible**, or **admissible for** $F$, iff, for every $n$-ary letter $L$ of $F$, the following two conditions are satisfied:

**(i)** $L^*$ does not depend on any variables that are not in its canonical tuple but occur in $F$.

**(ii)** Suppose $F$ contains an occurrence of $L(t_1, \ldots, t_n)$, the occurrence of $t_i$ ($1 \leq i \leq n$) within which is blindly bound in $F$. Then $L^*$ is unistructural in the $i$th variable of its canonical tuple.

Notice that condition (ii) of the above definition is automatically satisfied when $L$ is elementary, because an elementary problem is always unistructural. That is why this condition was absent in [6]. In most typical cases we will be interested in interpretations $^*$ that interpret each letter $L$ as a finitary unistructural game $L^*(x_1, \ldots, x_n)$ which does not depend on any variables other than $x_1, \ldots, x_n$, so that both conditions (i) and (ii) of Definition 2.1 will be automatically satisfied. With this remark in mind, henceforth we may sometimes omit "$F$-admissible" and simply say "interpretation"; every time an expression $F^*$ is used in a context, it should be understood that the range of $^*$ is restricted to $F$-admissible interpretations.

---

[2] A more elaborated discussion of applied systems based on computability logic can be found in Section 8 of [3] or Section 10 of [8]. See also Sections 26-28 of [2].

Every interpretation $^*$ extends from letters to formulas in the obvious way: where $L$ is an $n$-ary letter interpreted as $L^*(x_1, \ldots, x_n)$ and $t_1, \ldots, t_n$ are any terms, $\bigl(L(t_1, \ldots, t_n)\bigr)^* = L^*(t_1, \ldots, t_n)$; $\top^* = \top$; $(\neg G)^* = \neg(G^*)$; $(G_1 \sqcap \ldots \sqcap G_n)^* = G_1^* \sqcap \ldots \sqcap G_n^*$; $(\forall x G)^* = \forall x(G^*)$; etc. We say that a formula $F$ is **valid** iff, for every $F$-admissible interpretation $^*$, the problem $F^*$ is computable.

Since the blind operations are only partial functions, the above inductive definition of $F^*$ does not quite automatically guarantee that $F^*$ is always defined, so the following fact needs to be verified for safety, which will be formally done in Section 7:

**Fact 2.2** *For any formula $F$ and $F$-admissible interpretation $^*$, the game $F^*$ is defined.*

We extend to the language of **CL4** the notational conventions from [6] (Section 5) regarding the meaning of $E[x_1/t_1, \ldots, x_n/t_n]$ and representing formulas in the form $E(x_1, \ldots, x_n)$, which sets a context allowing us to write $E(t_1, \ldots, t_n)$ instead of $E[x_1/t_1, \ldots, x_n/t_n]$.

The terms "negative occurrence" and "positive occurrence" have the same meaning as before. Similarly, a **surface occurrence** of a subexpression in a formula is an occurrence that is not in the scope of $\sqcap, \sqcup, \sqcap, \sqcup$. When a formula does not contain any of these four operators, it is said to be **elementary**. The **elementarization** of a formula $E$ is the result of replacing in $E$ every surface occurrence of the form $G_1 \sqcap \ldots \sqcap G_n$ or $\sqcap x G$ by $\top$, every surface occurrence of the form $G_1 \sqcup \ldots \sqcup G_n$ or $\sqcup x G$ by $\bot$, every positive surface occurrence of each general atom by $\bot$, and every negative surface occurrence of each general atom by $\top$. A formula is **stable** iff its elementarization is a valid formula of classical logic.[3] Otherwise it is **instable**.

The rules of inference of **CL4** are the three rules **A**, **B1** and **B2** of **CL3** — only, now applied to any **CL4**-formulas rather than just **CL3**-formulas — plus one single additional rule **C**. Here are all four rules:

**A** $\vec{H} \mapsto E$, where $E$ is stable and $\vec{H}$ is a set of formulas satisfying the following conditions:

(i) Whenever $E$ has a positive (resp. negative) surface occurrence of a subformula $G_1 \sqcap \ldots \sqcap G_n$ (resp. $G_1 \sqcup \ldots \sqcup G_n$), for each $i \in \{1, \ldots, n\}$, $\vec{H}$ contains the result of replacing that occurrence in $E$ by $G_i$;

(ii) Whenever $E$ has a positive (resp. negative) surface occurrence of a subformula $\sqcap x G(x)$ (resp. $\sqcup x G(x)$), $\vec{H}$ contains the result of replacing that occurrence in $E$ by $G(y)$ for some variable $y$ not occurring in $E$.

**B1** $H \mapsto E$, where $H$ is the result of replacing in $E$ a negative (resp. positive) surface occurrence of a subformula $G_1 \sqcap \ldots \sqcap G_n$ (resp. $G_1 \sqcup \ldots \sqcup G_n$) by $G_i$ for some $i \in \{1, \ldots, n\}$.

**B2** $H \mapsto E$, where $H$ is the result of replacing in $E$ a negative (resp. positive) surface occurrence of a subformula $\sqcap x G(x)$ (resp. $\sqcup x G(x)$) by $G(t)$ for some term $t$ such that (if $t$ is a variable) neither the above occurrence of $\sqcap x G(x)$ (resp. $\sqcup x G(x)$) in $E$ nor any of the free occurrences of $x$ in $G(x)$ are in the scope of $\forall t, \exists t, \sqcap t$ or $\sqcup t$.

**C** $H \mapsto E$, where $H$ is the result of replacing in $E$ two — one positive and one negative — surface occurrences of some $n$-ary general letter by an $n$-ary non-logical elementary letter that does not occur in $E$.

Let us agree that, throughout this paper (with one little exception in Section 5), the uppercase $P, Q, R, S$ always stand for general letters and the lowercase $p, q, r, s$ for non-logical elementary letters. In any given context, all these letters will be assumed to be pairwise distinct, and the arity of each letter will be the length of the tuple of terms attached to it. When a letter $L$ is 0-ary, we write $L$ rather than $L()$. $x, y, z$ will always stand for variables (again, usually assumed to be pairwise distinct), $t$ for terms, and $c$ for constants.

---

[3] Stability is meant to be a syntactic concept, yet its definition seemingly relies on the semantical notion of classical validity. No problem: in view of Gödel's completeness theorem, "valid formula of classical logic" can be understood as a lazy way to say "provable formula of classical predicate calculus".

Looking at a few examples should help us get a syntactic feel of our most unusual deductive system. The following is a **CL4**-proof of $\sqcap x \sqcup y (P(x) \to P(y))$:

1. $p(z) \to p(z)$ (from {} by Rule **A**)
2. $P(z) \to P(z)$ (from 1 by Rule **C**)
3. $\sqcup y (P(z) \to P(y))$ (from 2 by Rule **B2**)
4. $\sqcap x \sqcup y (P(x) \to P(y))$ (from {3} by Rule **A**)

On the other hand, $\mathbf{CL4} \not\vdash \sqcup y \sqcap x (P(x) \to P(y))$. Indeed, obviously this instable formula cannot be the conclusion of any rule but **B2**. If it is derived by this rule, the premise should be $\sqcap x (P(x) \to P(t))$ for some term $t$ different from $x$. $\sqcap x (P(x) \to P(t))$, in turn, could only be derived by Rule **A** where, for some variable $y$ different from $t$, $P(y) \to P(t)$ is a (the) premise. The latter is an instable formula and does not contain choice operators, so the only rule by which it can be derived is **C**, where the premise is $p(y) \to p(t)$ for some elementary letter $p$. Now we deal with a classically non-valid and hence instable elementary formula, and it cannot be derived by any of the four rules of **CL4**.

Note that, in contrast, the "blind version" $\exists y \forall x (P(x) \to P(y))$ of $\sqcup y \sqcap x (P(x) \to P(y))$ is provable:

1. $\exists y \forall x (p(x) \to p(y))$ (from {} by Rule **A**)
2. $\exists y \forall x (P(x) \to P(y))$ (from 1 by Rule **C**)

'*There is $y$ such that, for all $x$, $P(x) \to P(y)$*' is true yet not in a constructive sense, thus belonging to the kind of principles that have been fueling those endless and fruitless fights between the classically- and constructivistically-minded. Computability logic is offering a peaceful settlement, telling the arguing parties: "There is no need to fight at all. It appears that you simply have two different concepts of *there is/for all*. So, why not also use two different names: $\exists/\forall$ and $\sqcup/\sqcap$. Yes, $\exists y \forall x (P(x) \to P(y))$ is indeed right; and yes, $\sqcup y \sqcap x (P(x) \to P(y))$ is indeed wrong." Clauses 1 and 2 of Exercise 2.3 illustrate a similar solution for *excluded middle* — the most typically attacked principle of classical logic.

The above-said remains true with $p$ instead of $P$, for what was relevant there was the difference between the constructive and non-constructive versions of logical operators rather than how atoms were understood. Then how about the difference between the elementary and non-elementary versions of atoms? This distinction allows computability logic to again act in its noble role of a reconciliator/integrator, but this time between classical and linear logics, telling them: "It appears that you have two different concepts of the objects that logic is meant to study. So, why not also use two different sorts of atoms to represent such objects: elementary atoms $p, q, \ldots$, and general atoms $P, Q, \ldots$. Yes, $p \to p \wedge p$ is indeed right; and yes, $P \to P \wedge P$ (Exercise 2.3(4)) is indeed wrong". However, the term "linear logic" in this context should be understood in a very generous sense, referring not to the particular deductive system **LL** proposed by Girard but rather to the general philosophy and intuitions traditionally associated with **LL** and some other, earlier-but less-known, substructural logics. For, as pointed out in [2, 6, 3, 7, 8], computability logic considers **LL** — with its multiplicatives seen as parallel operators and additives as choice operators — incomplete, only partially agreeing with **LL** regarding what principles should be deemed wrong. An example of such a principle is $P \wedge P \to P$ (Exercise 2.3(3)), accepted by **CL4** but rejected by **LL**. This formula becomes derivable if the weakening rule is added to **LL**, i.e. in the substructural logic in the past known under the name **BCK**, and after the conception of **LL** more frequently referred to as the additive-multiplicative fragment of *affine logic*. But just adding the weakening rule to **LL** is not sufficient to save the case. There are **CL4**-provable formulas derivable in neither **LL** nor **BCK**.[4] Clause 16 of Exercise 2.3 provides an example. Another example is Blass's [1] principle $(P \wedge Q) \vee (R \wedge S) \to (P \vee R) \wedge (Q \vee S)$.

**Exercise 2.3** In clauses 12 and 13 below, "$\mathbf{CL4} \vdash E \Leftrightarrow F$" stands for "$\mathbf{CL4} \vdash E \to F$ *and* $\mathbf{CL4} \vdash F \to E$". Verify that:

1. $\mathbf{CL4} \vdash P \vee \neg P$.
2. $\mathbf{CL4} \not\vdash P \sqcup \neg P$. Compare with 1.
3. $\mathbf{CL4} \vdash P \wedge P \to P$.

---

[4]While **BCK** is incomplete and hence so is affine logic as a conservatively extension of it, affine logic in its full additive-multiplicative-exponential language has been proven ([8]) to be complete with respect to the semantics of computability logic.

4. **CL4** $\not\vdash P \to P \wedge P$. Compare with 3,5.
5. **CL4** $\vdash P \to P \sqcap P$.
6. **CL4** $\vdash (P \sqcup Q) \wedge (P \sqcup R) \to P \sqcup (Q \wedge R)$. Remember the Go/Chess/Checkers example from Section 1.
7. **CL4** $\not\vdash P \sqcup (Q \wedge R) \to (P \sqcup Q) \wedge (P \sqcup R)$. Compare with 6,8.
8. **CL4** $\vdash p \sqcup (Q \wedge R) \to (p \sqcup Q) \wedge (p \sqcup R)$.
9. **CL4** $\not\vdash p \sqcap (Q \wedge R) \to (p \sqcap Q) \wedge (p \sqcap R)$. Compare with 8.
10. **CL4** $\vdash \forall x P(x) \to \sqcap x P(x)$.
11. **CL4** $\not\vdash \sqcap x P(x) \to \forall x P(x)$. Compare with 10.
12. **CL4** $\vdash \exists x P(x) \sqcap \exists x Q(x) \Leftrightarrow \exists x \bigl(P(x) \sqcap Q(x)\bigr)$. Similarly for $\sqcup$ instead of $\sqcap$, and/or $\forall$ instead of $\exists$.
13. **CL4** $\vdash \sqcap x \exists y P(x,y) \Leftrightarrow \exists y \sqcap x P(x,y)$. Similarly for $\sqcup$ instead of $\sqcap$, and/or $\forall$ instead of $\exists$.
14. **CL4** $\vdash \forall x \bigl(P(x) \wedge Q(x)\bigr) \to \forall x P(x) \wedge \forall x Q(x)$.
15. **CL4** $\not\vdash \sqcap x \bigl(P(x) \wedge Q(x)\bigr) \to \sqcap x P(x) \wedge \sqcap x Q(x)$. Compare with 14.
16. **CL4** $\vdash \sqcap x \Bigl(\bigl(P(x) \wedge \sqcap x Q(x)\bigr) \sqcap \bigl(\sqcap x P(x) \wedge Q(x)\bigr)\Bigr) \to \sqcap x P(x) \wedge \sqcap x Q(x)$.

Taking into account that classical validity and hence stability is recursively enumerable, the following fact is obvious:

**Fact 2.4** (*The set of theorems of*) **CL4** *is recursively enumerable.*

In Section 6 we will also prove that

**Theorem 2.5** *The $\forall, \exists$-free fragment of* (*the set of theorems of*) **CL4** *is decidable.*

The above theorem — just as the similar result for **CL3** known from [6] — establishes a nice fact, remarkably contrasting with the situation in classical logic: note that the $\forall, \exists$-free fragment of **CL4** is still first-order as it contains the quantifiers $\sqcap, \sqcup$. This fragment is also natural as it gets rid of the only operators of the language that produce games with imperfect information.

Next, based on the straightforward observation that elementary formulas are derivable in **CL4** (in particular, from the empty set of premises by Rule **A**) exactly when they are classically valid and hence derivable in classical predicate calculus, we have:

**Fact 2.6** **CL4** *is a conservative extension of classical predicate logic: the latter is nothing but the elementary fragment of the former.*

The following main Theorem 2.7, even though by an order of magnitude more informative than Gödel's completeness theorem for classical logic which it implies (by Fact 2.6) as a special case, is perhaps only one of the first steps on the way of in-depth study of computability logic. Seeing what happens if we add the recurrence group of operators (see Subsection 4.6 of [8]) to the language of **CL4** remains a challenging but worthy task for the future to pursue.

**Theorem 2.7** **CL4** $\vdash F$ *iff* $F$ *is valid* (*any formula $F$*). *Furthermore:*

**Uniform-Constructive Soundness:** *There is an effective procedure that takes a **CL4**-proof of an arbitrary formula $F$ and constructs an HPM that wins $F^*$ for every interpretation $*$.*

**Strong Completeness:** *If* **CL4** $\not\vdash F$, *then $F^*$ is not computable for some interpretation $*$ that interprets elementary letters as finitary predicates of arithmetical complexity $\Delta_2$, and interprets general letters as $\sqcap, \sqcup$-combinations of finitary predicates of arithmetical complexity $\Delta_2$.*

It was shown in [6] that the soundness/completeness result for **CL3** implied a positive verification of Conjectures 24.4, 25.4, 26.2 of [2] restricted to the language of that logic. For similar reasons, our Theorem 2.7 signifies a positive verification of those three conjectures restricted to the significantly more expressive language of **CL4**. The rest of this paper — for the exception of the last three short sections — is devoted to a proof of Theorem 2.7, which is nothing but a combination of Lemmas 4.1 (soundness) and 5.1 (completeness).

# 3 Preliminaries

## 3.1 From formulas to hyperformulas

In the bottom-up (from conclusion to premises) view, Rule **C** introduces two occurrences of some new non-logical elementary letter. For technical convenience, we want to differentiate elementary letters introduced this way from all other elementary letters, and also to somehow keep track of the exact origin of each such elementary letter $q$ — that is, remember what general letter $P$ was replaced by $q$ when Rule **C** was applied. For this purpose, we extend the language of **CL4** by adding to it a new sort of non-logical letters (correspondingly extending the scope of the common term "letter") called **hybrid**. In particular, each $n$-ary hybrid letter is a pair consisting of an $n$-ary general letter $P$, called its **general component**, and an $n$-ary non-logical elementary letter $q$, called its **elementary component**. We denote such a pair by $P_q$. As we are going to see later, the presence of $P_q$ in a (modified **CL4**-) proof will be an indication of the fact that, in the bottom-up view of proofs, $q$ has been introduced by Rule **C** and that when this happened, the general letter that $q$ replaced was $P$.

In this new context, an **atom** — in particular, $n$-ary atom — is the expression $L(t_1, \ldots, t_n)$, where $L$ is an $n$-ary (elementary, general or hybrid) letter and the $t_i$ are any terms. We will often abbreviate such a tuple $t_1, \ldots, t_n$ as $\vec{t}$. The atom $L(\vec{t})$ is said to be $L$-**based**, and any two $L$-based atoms are said to be **same-base**. An $L$-based atom is said to be elementary, general, hybrid, logical or non-logical iff the letter $L$ is so. What we call **hyperformulas** are defined in the same way as formulas, with the only difference that now atomic expressions can be of any of the three (elementary, general or hybrid) sorts. "Subhyperformula" in the context of hyperformulas will mean the same as "subformula" in the context of formulas. Our conventions regarding the notation $E[x_1/t_1, \ldots, x_n/t_n]$, or representing formulas in the form $E(x_1, \ldots, x_n)$ and then writing $E(t_1, \ldots, t_n)$ for $E[x_1/t_1, \ldots, x_n/t_n]$, extend from formulas to hyperformulas in the obvious way. So do the concepts such as a *surface* (or *positive*, or *negative*) occurrence of a subexpression, a *free* (or *bound*, or *blindly bound*) occurrence of a variable, etc. The concept of a free occurrence naturally extends to all terms by stipulating that an occurrence of a constant in a hyperformula is always free. By a **free variable** of a given hyperformula $E$ we mean a variable that has at least one free occurrence in $E$. If there are no such variables, then $E$ is said to be **closed**.

A hyperformula $E$ is said to be **balanced** iff, for every hybrid letter $P_q$ occurring in $E$, the following conditions are satisfied:

1. $E$ has exactly two occurrences of $P_q$, where one occurrence is positive and the other occurrence is negative, and both occurrences are surface occurrences;

2. The elementary letter $q$ does not occur in $E$, nor is it the elementary component of any hybrid letter occurring in $E$ other than $P_q$.

An **elementary hyperformula** is one not containing $\sqcap, \sqcup, \sqcap$ and $\sqcup$, as well as general and hybrid atoms. Thus, elementary hyperformula and elementary formula mean the same. The **elementarization**

$$\|E\|$$

of a hyperformula $E$ is the result of replacing, in $E$, each surface occurrence of the form $G_1 \sqcap \ldots \sqcap G_n$ or $\sqcap xG$ by $\top$, each surface occurrence of the form $G_1 \sqcup \ldots \sqcup G_n$ or $\sqcup xG$ by $\bot$, every positive (resp. negative) surface occurrence of each general atom by $\bot$ (resp. $\top$), and every surface occurrence of each hybrid letter by the elementary component of that letter. As in the case of formulas, we say that a hyperformula $E$ is **stable** iff $\|E\|$ is valid in the classical sense; otherwise it is **instable**.

In our soundness proof for **CL4** we will employ a "version" of **CL4** called **CL4°**. Unlike **CL4** whose language consists only of formulas, the language of **CL4°** allows any balanced hyperformulas. The rules of **CL4°** are Rules **A**, **B1** and **B2** of **CL4** — only now applied to any balanced hyperformulas rather than just formulas — plus the following rule **C°** instead of **C**:

**C°**    $H \mapsto E$, where $E$ is the result of replacing in $H$ both occurrences of some hybrid letter $P_q$ by its general component $P$.

**Lemma 3.1** *For any formula $F$, if $\mathbf{CL4} \vdash F$, then $\mathbf{CL4°} \vdash F$. Furthermore, there is an effective procedure that converts any $\mathbf{CL4}$-proof of any formula $F$ into a $\mathbf{CL4°}$-proof of $F$.*

**Idea.** Every application of Rule **C** naturally turns into an application of Rule **C°**.

**Proof.** Consider any **CL4**-proof tree for $F$, i.e. a tree every node of which is labeled with a formula that follows by one of the rules of **CL4** from (the labels of) its children, with $F$ being the label of the root. By abuse of terminology, here we identify the nodes of this tree with their labels, even though, of course, it may be the case that different nodes have the same label. For each node $E$ of the tree that is derived from its child $H$ by Rule **C**, — in particular, where $H$ is the result of replacing in $E$ a positive and a negative surface occurrences of an $n$-ary general letter $P$ by an $n$-ary non-logical elementary letter $q$, — do the following: replace $q$ by the hybrid letter $P_q$ in $H$ as well as in all of its descendants in the tree. It is not hard to see that this way we will get a **CL4°**-proof of $F$.  □

Let $E$ be a balanced hyperformula, and $P_q$ an $n$-ary hybrid letter of $E$ with $P_q(t_1, \ldots, t_n)$ and $P_q(t'_1, \ldots, t'_n)$ being the two $P_q$-based atoms occurring in $E$. We say that $P_q$ is an **unreasonable** hybrid letter of $E$ iff, for some $i$ with $1 \leq i \leq n$, the occurrence of $t_i$ within $P_q(t_1, \ldots, t_n)$ and the occurrence of $t'_i$ within $P_q(t'_1, \ldots, t'_n)$ are both free in $E$, and $t_i$ and $t'_i$ are (graphically) different terms. We say that a hyperformula is **reasonable** iff it is balanced and does not have any unreasonable hybrid letters. And we say that a **CL4°**-proof is **reasonable** iff all of its hyperformulas are reasonable.

Intuitively, as will be seen shortly, the presence of unreasonable hyperformulas in a **CL4°**-proof signifies that some really "unreasonable" steps have been made in the proof. Specifically, in the bottom-up view, the applications of Rule **C°** that introduced unreasonable hybrid letters were unnecessary, as the premise of such an application is never any "more provable" (i.e. never has a shorter or simpler proof) than the conclusion.

**Lemma 3.2** *Assume $E$ is a balanced hyperformula and $P_q$ an unreasonable hybrid letter of $E$. Let $F$ be the result of replacing in $E$ both of the occurrences of letter $P_q$ by $P$. Then, if $E$ is stable, so is $F$.*

**Idea.** If $P_q$ is unreasonable, due to the presence of two different free terms in the two $q$-based atoms of $\|E\|$ such as, say, $q(1, \vec{x})$ and $q(2, \vec{y})$, these atoms can be treated as if they simply were atoms based on two different letters, such as $q_1(\vec{x})$ and $q_2(\vec{y})$. Then $q_1(\vec{x})$ and $q_2(\vec{y})$ are fully isolated — not only from everything else but also from each other — in the sense that each of the letters $q_1, q_2$ occurs only once in the formula. It is known that in a classically valid formula such isolated atoms can be replaced by whatever formulas without affecting validity. So, if $\|E\|$ is valid, it will remain so after replacing in it $q_1(\vec{x})$ by $\bot$ and $q_2(\vec{y})$ by $\top$ (or vice versa, depending on which one is positive and which one is negative). The result of such a replacement is exactly what $\|F\|$ is.

**Proof.** Assume the conditions of the lemma. For simplicity, we may also assume that $E$ is closed and hence so is $F$. Indeed, otherwise, for every variable $x$, we can replace all free occurrences of $x$ in these hyperformulas by some constant not occurring in them, with different variables replaced by different constants. Obviously this conversion would have no effect on being balanced, reasonable or stable.

Suppose $F$ is instable, i.e., for some classical model $M$,

$$\|F\| \text{ is false in } M. \tag{1}$$

Our goal is to show that then $E$ is instable, too. Let $n$ be the arity of $P_q$, and assume that the two — positive and negative — occurrences of $P_q$ in $E$ are within atoms $P_q(t_1, \ldots, t_n)$ and $P_q(t'_1, \ldots, t'_n)$, respectively. The unreasonably of $P_q$ means that, for some $i$ with $1 \leq i \leq n$, neither the occurrence of $t_i$ within $P_q(t_1, \ldots, t_n)$ nor the occurrence of $t'_i$ within $P_q(t'_1, \ldots, t'_n)$ is bound in $E$, and $t_i$ and $t'_i$ are different terms. Let us take a note that these two terms are constants because, as we agreed, $E$ has no free occurrences of variables.

Based on known facts from classical logic, we may assume that the universe of discourse of the above model $M$ is just the (countably infinite) set of our constants, and that each constant $c$ is interpreted as $c$ itself. Also, unlike $\|E\|$, $\|F\|$ does not contain $q$, and therefore (1) will continue to hold under arbitrary assumptions regarding how $q$ is interpreted in $M$. So, an assumption we are making is that, for any constants $c_1, \ldots, c_n$, $q(c_1, \ldots, c_{i-1}, t_i, c_{i+1}, \ldots, c_n)$ is false in $M$ and $q(c_1, \ldots, c_{i-1}, t'_i, c_{i+1}, \ldots, c_n)$ is true. Thus, in $M$, $q(t_1, \ldots, t_n)$ is equivalent (in the standard classical sense) to $\bot$ and $q(t'_1, \ldots, t'_n)$ equivalent to $\top$. Observe that $\|F\|$ is nothing but the result of replacing, in $\|E\|$, the occurrence of $q(t_1, \ldots, t_n)$ by $\bot$ and the occurrence of $q(t'_1, \ldots, t'_n)$ by $\top$. Replacing subformulas by equivalent ones does not change the truth status of a formula. So $\|E\|$ has the same truth value in $M$ as $\|F\|$ does, i.e., by (1), is false. This means that $E$ is instable.  □

**Lemma 3.3** *For any reasonable hyperformula $F$, if $\mathbf{CL4}^\circ \vdash F$, then $F$ has a reasonable $\mathbf{CL4}^\circ$-proof. Furthermore, there is an effective procedure that converts any $\mathbf{CL4}^\circ$-proof of any reasonable hyperformula $F$ into a reasonable $\mathbf{CL4}^\circ$-proof of $F$.*

**Idea.** Given a $\mathbf{CL4}^\circ$-proof of a reasonable hyperformula, mechanically disregard in it all of the "unreasonable" applications of Rule $\mathbf{C}^\circ$, i.e. the applications that (in the bottom-up view) introduced unreasonable letters.

**Proof.** In the present context we prefer to see proofs as sequences rather than trees of hyperformulas. For any balanced hyperformula $E$, let $\tilde{E}$ mean the result of replacing in $E$ every (occurrence of) every unreasonable hybrid letter by the general component of that letter.

Consider a $\mathbf{CL4}^\circ$-proof $\mathcal{P}$ of a reasonable hyperformula $F$. We need to show how to (effectively) convert $\mathcal{P}$ into a reasonable $\mathbf{CL4}^\circ$-proof of $F$. Nothing is easier than to do such a conversion: just replace in $\mathcal{P}$ every hyperformula $E$ by $\tilde{E}$. We claim that the resulting sequence $\tilde{\mathcal{P}}$ will be a reasonable $\mathbf{CL4}^\circ$-proof of $F$.

Such a claim can be verified by showing that every hyperformula $\tilde{E}$ of $\tilde{\mathcal{P}}$ follows by one of the rules of $\mathbf{CL4}^\circ$ from a (possibly empty) set of earlier hyperformulas of $\tilde{\mathcal{P}}$. Indeed, assume $E$ is derived in $\mathcal{P}$ from $\{H_1, \ldots, H_n\}$ by Rule $\mathbf{A}$. This means that $E$ is stable. Applying Lemma 3.2 as many times as the number of unreasonable hybrid letters in $E$, we find that $\tilde{E}$ is also stable. Then, obviously, $\tilde{E}$ follows from $\{\tilde{H}_1, \ldots, \tilde{H}_n\}$ by Rule $\mathbf{A}$. In case $E$ was derived in $\mathcal{P}$ from a premise $H$ by Rule $\mathbf{B1}$ or $\mathbf{B2}$, evidently we again have that $\tilde{E}$ follows from $\tilde{H}$ by the same rule. Suppose now $E$ is derived in $\mathcal{P}$ from $H$ by Rule $\mathbf{C}^\circ$ — in particular, $H$ is the result of replacing in $E$ two (positive and negative) surface occurrences of some general letter $P$ by a hybrid letter $P_q$. If $P_q$ is not an unreasonable hybrid letter of $H$, it is obvious that $\tilde{E}$ follows from $\tilde{H}$ by Rule $\mathbf{C}^\circ$. Suppose now $P_q$ is an unreasonable hybrid letter of $H$. Notice that then $\tilde{E} = \tilde{H}$; since $H$ occurs in $\mathcal{P}$ earlier than $E$ does, we may assume (by the induction hypothesis) that $\tilde{H}$ follows from some earlier hyperformulas of $\tilde{\mathcal{P}}$ by one of the rules. Hence, so does $\tilde{E}$. □

## 3.2 Interpretations of hyperformulas

By the **general dehybridization** of a hyperformula $F$ we mean the formula that is the result of replacing in $F$ every hybrid letter by the general component of that letter. Where $F$ is a hyperformula and $G$ its general dehybridization, we say that an interpretation $*$ is $F$-**admissible** iff it is $G$-admissible, and we define the game $F^*$ to be $G^*$. By the earlier definition, remembering that "problem"="static game", every interpretation $*$ sends letters to static games, and it is known from [2] (Theorem 14.1) that all of our game operations preserve the static property of games. So, the game $F^*$ (any hyperformula $F$, any $F$-admissible interpretation $*$) is always static. We will often implicitly rely on this fact in the present paper.

We say that two games $A$ and $B$ are **equistructural** iff $\mathbf{Lr}_e^A = \mathbf{Lr}_e^B$ for every valuation $e$.

**Lemma 3.4** *Assume $F$ is a reasonable hyperformula containing an $n$-ary hybrid letter $P_q$, $*$ is an $F$-admissible interpretation that interprets $P$ as $P^*(x_1, \ldots, x_n)$, and $P_q(t_1, \ldots, t_n)$ and $P_q(t'_1, \ldots, t'_n)$ are the two $P_q$-based atoms occurring in $F$. Then the games $P^*(t_1, \ldots, t_n)$ and $P^*(t'_1, \ldots, t'_n)$ are equistructural.*

**Idea.** Due to reasonability, $P^*(t_1, \ldots, t_n)$ and $P^*(t'_1, \ldots, t'_n)$ only differ from each other in terms that are blindly bound (in at least one of the atoms). Condition (ii) of admissibility then guarantees that the structures of the above two games (their $\mathbf{Lr}$ components) do not depend on the values of such terms.

**Proof.** Assume the conditions of the lemma. Without loss of generality and for convenience of representation, let us also assume that, for some $i$ (fix it) with $0 \leq i \leq n$, we have $t_1 \neq t'_1$, ..., $t_i \neq t'_i$ and $t_{i+1} = t'_{i+1}$, ..., $t_n = t'_n$. So, we can rewrite the two $P_q$-based atoms of $F$ as $P_q(t_1, \ldots, t_i, t_{i+1}, \ldots, t_n)$ and $P_q(t'_1, \ldots, t'_i, t_{i+1}, \ldots, t_n)$. The reasonabity of $F$ implies that, for each $j$ with $1 \leq j \leq i$, either the occurrence of $t_j$ within $P_q(t_1, \ldots, t_i, t_{i+1}, \ldots, t_n)$ or the occurrence of $t'_j$ within $P_q(t'_1, \ldots, t'_i, t_{i+1}, \ldots, t_n)$ is blindly bound in $F$ and hence in its general dehybridization. In either case, by clause (ii) of Definition 2.1, $P^*(x_1, \ldots, x_i, x_{i+1}, \ldots, x_n)$ is unistructural in $x_j$. Thus,

$$P^*(x_1, \ldots, x_i, x_{i+1}, \ldots, x_n) \text{ is unistructural in each of the variables } x_1, \ldots, x_i. \tag{2}$$

Pick any valuation $e$. Looking back at Definition 4.1 of [6], we find that $\mathbf{Lr}_e^{P^*(t_1,\ldots,t_i,t_{i+1},\ldots,t_n)} = \mathbf{Lr}_g^{P^*(x_1,\ldots,x_i,x_{i+1},\ldots,x_n)}$ and $\mathbf{Lr}_e^{P^*(t_1',\ldots,t_i',t_{i+1},\ldots,t_n)} = \mathbf{Lr}_{g'}^{P^*(x_1,\ldots,x_i,x_{i+1},\ldots,x_n)}$, where $g$ and $g'$ are certain valuations that agree with $e$ on all variables that are not among $x_1,\ldots,x_n$, and agree with each other on $x_{i+1},\ldots,x_n$. Thus, the only variables on which $g$ and $g'$ disagree can be $x_1,\ldots,x_i$. But, by (2), $\mathbf{Lr}_g^{P^*(x_1,\ldots,x_n)}$ does not depend on the value of $g$ at $x_1,\ldots,x_i$. Hence $\mathbf{Lr}_g^{P^*(x_1,\ldots,x_n)} = \mathbf{Lr}_{g'}^{P^*(x_1,\ldots,x_n)}$, and thus $\mathbf{Lr}_e^{P^*(t_1,\ldots,t_n)} = \mathbf{Lr}_e^{P^*(t_1',\ldots,t_n')}$. Since $e$ was arbitrary, we conclude that $P^*(t_1,\ldots,t_n)$ and $P^*(t_1',\ldots,t_n')$ are equistructural. □

The concept of a perfect interpretation introduced in [6] naturally extends to the language of **CL4**: an interpretation $*$ is **perfect** iff, for every (elementary or general) letter $L$, $L^*$ is a finitary game that does not depend on any variables other than those from its canonical tuple. Next, for an interpretation $*$ and valuation $e$, the **perfect interpretation induced by** $(*,e)$ is the interpretation $\star$ that interprets every $n$-ary (elementary or general) letter $L$ as $L^\star(x_1,\ldots,x_n)$, where $x_1,\ldots,x_n$ is the canonical tuple of $L^*$ and $L^\star(x_1,\ldots,x_n)$ is the game such that, for every tuple $c_1,\ldots,c_n$ of constants, $L^\star(c_1,\ldots,c_n) = e[L^*(c_1,\ldots,c_n)]$. Intuitively, $L^\star$ is the "strictly $n$-ary version" of $L^*$, where the values of all variables that are not in the canonical tuple are fixed to the constants assigned to those variables by $e$, so that $L^\star$, unlike $L^*$, does not depend on any variables other than those from its canonical tuple, and $\star$ really *is* a perfect interpretation.

Many of the other concepts and notational conventions of [6] extend to our present context in an even more straightforward way and there is no need to officially redefine them. These include the notation $fF$ meaning the result of replacing in (hyper)formula $F$ every free occurrence of every variable by the constant which valuation $f$ assigns to that variable.

The following three Lemmas 3.5, 3.6 and 3.7 can be proven in the same way as Lemmas 7.2, 7.3 and 7.7 of [6]: the fact that those dealt with **CL3**-formulas rather than hyperformulas is hardly of any relevance.

**Lemma 3.5** *Suppose $x$ is a variable occurring in a hyperformula $F$. Then, for any ($F$-admissible) interpretation $*$, constant $c$ and subhyperformula $G$ of $F$, $(G[x/c])^* = G^*[x/c]$.*

**Lemma 3.6** *For any hyperformula $F$ and ($F$-admissible) perfect interpretation $*$, the game $F^*$ is (finitary and) does not depend on any variables that do not occur free in $F$; hence, if $F$ is closed, $F^*$ is a constant game.*

**Lemma 3.7** *For any hyperformula $F$, ($F$-admissible) interpretation $*$, and valuations $e$ and $f$ that agree on all free variables of $F$, we have $e[F^*] = e[(fF)^*]$.*

As we did in [6], with Lemma 3.6 in mind and in accordance with our conventions, as long as a hyperformula $F$ is closed and an interpretation $*$ is perfect, we can always safely omit the irrelevant valuation parameter $e$ in $\mathbf{Wn}_e^{F^*}$, $\mathbf{Lr}_e^{F^*}$ or $e[F^*]$, and simply write $\mathbf{Wn}^{F^*}$, $\mathbf{Lr}^{F^*}$ or $F^*$.

The following fact, on which we often implicitly rely in this paper, was straightforward in the context of **CL3**, but perhaps not quite so in our new context. A formal verification for it is given later in Section 8.

**Fact 3.8** *For any hyperformula $F$ and valuation $e$, whenever an interpretation $*$ is $F$-admissible, so is the perfect interpretation $\star$ induced by $(*,e)$.*

**Lemma 3.9** *Assume $F$ is a closed hyperformula, $e$ any valuation, $*$ any $F$-admissible interpretation, and $\star$ the perfect interpretation induced by $(*,e)$. Then $e[F^*] = F^\star$.*

**Idea.** When $F$ is a closed atom, the effect $e[F^*] = F^\star$ is the very meaning of the perfect interpretation $\star$ induced by $(*,e)$. The same effect seamlessly extends from atoms to compound closed (hyper)formulas.

**Proof.** Let $F$, $e$, $*$, $\star$ be as above. We may assume that $F$ does not contain hybrid atoms, for if it does, replace $F$ by its general dehybridization. Of course, $*$ remains admissible for every $G$ that is a subformula of $F$ or the result of substituting in such a subformula some free occurrences of variables by constants. We prove the lemma by induction on the complexity of $F$. If $F$ is an atom, $e[F^*] = F^\star$ is immediate from the definition of the perfect interpretation induced by $(*,e)$. And the cases when $F$ is non-atomic with its main operator being among $\neg, \wedge, \vee, \rightarrow, \sqcap, \sqcup$ are also straightforward because $e[\ldots]$, $^*$, $^\star$ commute with each of these operators. Finally, let $F = \mathcal{Q}xG$, where $x$ is a variable and $\mathcal{Q}$ a quantifier. Consider an arbitrary constant $c$. By Lemma 3.5, $G^*[x/c] = (G[x/c])^*$ and thus $e[G^*[x/c]] = e[(G[x/c])^*]$. But, by the induction hypothesis, $e[(G[x/c])^*] = (G[x/c])^\star$. Hence $e[G^*[x/c]] = (G[x/c])^\star$. We can rewrite the same as $e[G^*[x/c]] = e[(G[x/c])^\star]$ because, by Lemma 3.6, $(G[x/c])^\star$ is a constant game. Again by Lemma 3.5, $(G[x/c])^\star = G^\star[x/c]$, so that $e[G^*[x/c]] = e[G^\star[x/c]]$. Thus, for every constant $c$, $e[G^*[x/c]]$ and $e[G^\star[x/c]]$ are the same. Now, examining the definitions of $\forall, \exists, \sqcap, \sqcup$ (Section 4 of [6]) and seeing that $e[\mathcal{Q}xA]$ only depends on $e[\mathcal{Q}xA[x/c]]$ for $c \in \{0, 1, \ldots\}$, we find $e[\mathcal{Q}xG^*] = e[\mathcal{Q}xG^\star]$, i.e. $e[F^*] = e[F^\star]$. Again, since $F^\star$ is a constant game, $e[F^\star]$ can be rewritten as $F^\star$, concluding that $e[F^*] = F^\star$. □

## 3.3 Prefixation lemmas

For a game $A$, we will be using the expression $\mathbf{LR}^A$ to denote the set of all **unilegal runs** of $A$, i.e.

$$\mathbf{LR}^A = \{\Gamma \mid \Gamma \in \mathbf{Lr}_e^A \text{ for every valualuation } e\}.$$

Remember the notations $\Gamma^\gamma$, $\neg \Gamma$ and $\langle \Phi \rangle A$ (the $\Phi$-*prefixation* of $A$) from [6]. The game $\langle \Phi \rangle A$ — which in this paper we may sometimes also write just as $\Phi A$ — is defined if and only if $\Phi \in \mathbf{LR}^A$. For readability and compactness of formulations, let us agree that:

**Convention 3.10** Every time we write $\langle \Phi \rangle A$, we implicitly claim that this game is defined, i.e. $\Phi \in \mathbf{LR}^A$.

**Lemma 3.11** *Let $A, A_1, \ldots, A_n$ ($n \geq 2$) be any games, $x$ any variable, and $\Phi$ any position. For each of the following clauses, we (explicitly) assume that $\Phi$ is a unilegal position of the game to which $\Phi$-prefixation is applied; for clause 3 we also assume that $A$ is $x$-unistructural. Then:*
  1. $\langle \Phi \rangle \neg A = \neg(\langle \neg \Phi \rangle A)$.
  2. $\langle \Phi \rangle (A_1 \vee \ldots \vee A_n) = \langle \Phi^{1.} \rangle A_1 \vee \ldots \vee \langle \Phi^{n.} \rangle A_n$.
  3. $\langle \Phi \rangle \exists x A = \exists x \langle \Phi \rangle A$.

**Proof.** Clauses 1, 2 and 3 of this lemma can be easily proven by induction on the length of $\Phi$ based on clauses 1, 3 and 10 of Lemma 4.7 of [6], respectively, with inductive steps relying on the obvious fact that $\langle \Phi, \wp\alpha \rangle B = \langle \wp\alpha \rangle (\langle \Phi \rangle B)$ (see the footnote on page 24). Officially, clauses 1 and 2 have been verified in [5] (Lemmas 3.5 and 3.6), and the fact stated in clause 3 has been observed in Section 11 of [2] (equation (8) and the subsequent remark). □

By a **choice hyperformula** we mean a non-atomic hyperformula whose main operator is $\sqcap, \sqcup, \sqcap$ or $\sqcup$; depending on that operator, we may more specifically refer to such a hyperformula as a $\sqcap$- $\sqcup$-, $\sqcap$- or $\sqcup$-*hyperformula*. Henceforth we usually only deal with balanced (in fact, reasonable) hyperformulas, and their subhyperformulas that are choice hyperformulas are always simply formulas. However, for uniformity, we will still typically say "choice ($\sqcap$-, etc.) hyperformula" rather than "choice ($\sqcap$-, etc.) formula".

A **quasiatom** of a hyperformula $E$ is a surface occurrence of a subhyperformula in $E$ that is either an atom (of any of the three sorts) or a choice hyperformula. Note that a quasiatom is not just a subhyperformula but rather a subhyperformula together with a particular occurrence. E.g., in $P \wedge P$, the two different occurrences of $P$ present two different quasiatoms. However, for readability (and by abuse of concepts), we will often identify a quasiatom with the corresponding hyperformula $G$, and simply say "the quasiatom $G$" once it is clear from the context which of the possibly many occurrences of $G$ we mean. A quasiatom $G$ of a hyperformula $E$ is said to be *positive* (resp. *negative*) iff its occurrence in $E$ is positive (resp. negative). Similarly, such a quasiatom $G$ is *elementary*, *general*, *hybrid* or *choice* iff it is an elementary atom, general atom, hybrid atom, or choice hyperformula, respectively. Likewise for other terms such as *n-ary*, *L-based*, *same-base* etc. A *non-elementary* (*non-choice*, etc.) quasiatom means a quasiatom that is not an elementary (choice, etc.) one.

In Section 7 of [6] we defined the noun "$E$-**specification**" and the corresponding verb "to $E$-**specify**", with $E$ being a **CL3**-formula, specified objects being surface occurrences of its subformulas, and their $E$-specifications being strings acting as sorts of addresses of those occurrences/subformulas in the parse tree for the $(\vee, \wedge, \rightarrow)$-structure of $E$. This terminology straightforwardly extends to our new context where $E$ can be any hyperformula and $E$-specified objects be any surface occurrences of its subhyperformulas, including quasiatoms. Further extending it from quasiatoms to letters, the $E$-specification of a given surface occurrence of a letter $L$ means the $E$-specification of the (non-choice) quasiatom in which that occurrence of $L$ happens to be. For a hyperformula or letter $G$, we can say "the occurrence $\gamma$ of $G$ in $E$" to mean the surface occurrence of $G$ in $E$ that is $E$-specified by $\gamma$. Note that while a given string $\gamma$ can $E$-specify the surface occurrence of more than one subhyperformula, the quasiatom or letter $E$-specified by $\gamma$ is always unique. E.g., when $E = Q \wedge \neg P(x)$, two subformulas and one letter of $E$ are $E$-specified by '2.': $\neg P$, $P(x)$ and $P$; however, out of these three expressions, only the second one is a quasiatom and only the third one is a letter. So, quasiatoms and surface occurrences of letters of a given hyperformula $E$ can be uniquely identified by their $E$-specifications.

Where $\Gamma$ is a run, $E$ a hyperformula, $G$ a quasiatom of $E$ and $\gamma$ its $E$-specification, we define $\Gamma_E^\gamma$ by:

$$\Gamma_E^\gamma = \begin{cases} \Gamma^\gamma & \text{if } G \text{ is positive in } E; \\ \neg \Gamma^\gamma & \text{if } G \text{ is negative in } E. \end{cases}$$

By the **surface complexity** of a hyperformula $E$ we mean the number of surface occurrences of $\neg$, $\wedge$, $\vee$, $\rightarrow$, $\forall$, $\exists$ in $E$. A couple of forthcoming lemmas will be proven by induction on surface complexity. In the inductive steps of such proofs, we will only consider the cases when the main operator is $\neg$, $\vee$ or $\exists$. The cases with $\wedge$, $\rightarrow$ and $\forall$ can be safely omitted, as these three operators can be considered standard abbreviations in terms of $\neg$, $\vee$ and $\exists$.

According to our earlier conventions from [6], below and elsewhere $\wp$ ranges over $\{\top, \bot\}$.

**Lemma 3.12** *For any hyperformula $E$, $E$-admissible interpretation $*$ and run $\Gamma$, $\Gamma \in \mathbf{LR}^{E^*}$ iff every labmove of $\Gamma$ has the form $\wp\gamma\beta$ for some $\gamma$ that $E$-specifies a non-elementary quasiatom $G$ of $E$, such that $\Gamma_E^\gamma \in \mathbf{LR}^{G^*}$.*

**Idea.** The meaning of the above lemma is that a unilegal run of the game represented by a hyperformula $E$ consists of unilegal runs of the games represented by the quasiatoms of $E$, with the roles of the two players switched in negative quasiatoms. This is in concordance with the intuition — explained in [6] — that (uni)legal runs of $(\neg, \vee, \wedge, \rightarrow, \exists, \forall)$-combinations of games are nothing but interspersed (uni)legal runs of the component games, with $\neg$ and $\rightarrow$ (in its antecedent) switching players' roles. The lemma only talks about non-elementary quasiatoms, because elementary quasiatoms represent games without legal moves, and a legal run of a compound game would never include any moves made in elementary components of the game.

**Proof.** We prove Lemma 3.12 by induction on the surface complexity of $E$. For the basis of induction, assume $E$ is a quasiatom. If $E$ is an elementary quasiatom and hence $E^*$ is an elementary game, then $\Gamma \in \mathbf{LR}^{E^*}$ iff $\Gamma = \langle\rangle$, because, as we remember, $\langle\rangle$ is the only legal run of elementary games. And, since in this case $E$ has no non-elementary quasiatoms, the statement of the lemma is vacuously true. Suppose now $E$ is a non-elementary quasiatom. The occurrence of $E$ in itself is $E$-specified by the empty string $\epsilon$. Inserting $\epsilon$ does not change a string, so every labmove $\wp\beta$ of $\Gamma$ has the form $\wp\epsilon\beta$. And, of course, $\Gamma_E^\epsilon = \Gamma$. Therefore, again, what the lemma claims is trivially true.

For the inductive step, assume $E = \neg K$. By the definition of $\neg$, $\Gamma \in \mathbf{LR}^{E^*}$ iff $\neg\Gamma \in \mathbf{LR}^{K^*}$. In turn, by the induction hypothesis, $\neg\Gamma \in \mathbf{LR}^{K^*}$ iff every labmove of $\neg\Gamma$ has the form $\wp\gamma\beta$ for some $\gamma$ that $K$-specifies a non-elementary quasiatom $G$ of $K$, such that $(\neg\Gamma)_K^\gamma \in \mathbf{LR}^{G^*}$. But notice that the same $\gamma$ also $E$-specifies the same quasiatom $G$, and that $(\neg\Gamma)_K^\gamma = \Gamma_E^\gamma$. Hence, the statement of the lemma is correct.

Next, assume $E = K_1 \vee \ldots \vee K_n$. By the definition of $\vee$, $\Gamma \in \mathbf{LR}^{E^*}$ iff every labmove of $\Gamma$ has the form $\wp i.\alpha$ for some $i \in \{1, \ldots, n\}$ and, for each such $i$, $\Gamma^{i.} \in \mathbf{LR}^{K_i^*}$. In turn, by the induction hypothesis, $\Gamma^{i.} \in \mathbf{LR}^{K_i^*}$ iff every labmove of $\Gamma^{i.}$ has the form $\wp\delta\beta$ for some $\delta$ that $K_i$-specifies a non-elementary quasiatom $G$ of $K_i$, such that $(\Gamma^{i.})_{K_i}^\delta \in \mathbf{LR}^{G^*}$. Notice that the same $G$ is a quasiatom of $E$ which is $E$-specified by $i.\delta$, and that $(\Gamma^{i.})_{K_i}^\delta = \Gamma_E^{i.\delta}$. Thus, $\Gamma \in \mathbf{LR}^{E^*}$ iff every labmove of $\Gamma$ has the form $\wp i.\delta\beta$, where $i.\delta$ is the $E$-specification of a non-elementary quasiatom $G$ of $E$, such that $\Gamma_E^{i.\delta} \in \mathbf{LR}^{G^*}$. In other words, with $i.\delta$ in the role of $\gamma$, the statement of the lemma holds.

Finally, assume $E = \exists x K$. Taking into account that every quasiatom of $\exists x K$ is also a quasiatom of $K$ (and vice versa) with the same specification and same positive/negative status, this case is straightforward because, by the definition of $\exists$, $\mathbf{LR}^{\exists x K^*} = \mathbf{LR}^{K^*}$. □

For a run $\Gamma$ and string $\gamma$, we will be using the expression

$$\Gamma^{-\gamma}$$

to denote the result of deleting in $\Gamma$ every labmove of the form $\wp\gamma\beta$.

**Lemma 3.13** *Assume $E, F$ are hyperformulas, $G$ is a quasiatom of $E$, and $H$ is the result of replacing this quasiatom by $F$ in $E$. Further assume $\gamma$ is the $E$-specification of $G$, $*$ is an interpretation admissible for both $E$ and $H$, and $\Phi$ is a unilegal position of $E^*$ with $\langle\Phi_E^\gamma\rangle G^* = F^*$. Then $\langle\Phi\rangle E^* = \langle\Phi^{-\gamma}\rangle H^*$.*

**Idea.** Let us look at an example with $\Phi = \langle\bot 1.2, \top 2.1\rangle$. The effect of playing $\Phi$ in game $(A \sqcap B) \vee (C \sqcup D)$ is obviously $B \vee C$, in the sense that $\langle\Phi\rangle\big((A \sqcap B) \vee (C \sqcup D)\big) = B \vee C$. But the same effect is achieved by just playing $\langle\top 2.1\rangle$ in $B \vee (C \sqcup D)$. This is so because $\langle\top 2.1\rangle$ is the result of deleting from $\Phi$ the move(s) "meant" for the $A \sqcap B$ component of the game, and $B$ is the result of making those very moves in $A \sqcap B$. Thus, $\Phi$ can be replaced by $\langle\top 2.1\rangle$ provided that we also replace $A \sqcap B$ by $B$. One may guess that phenomena such as the just-observed $\langle\bot 1.2, \top 2.1\rangle\big((A \sqcap B) \vee (C \sqcup D)\big) = \langle\top 2.1\rangle\big(B \vee (C \sqcup D)\big)$ are no accident, and this is exactly what Lemma 3.13 asserts. In our example, identifying formulas with their interpretations, $E$ is $(A \sqcap B) \vee (C \sqcup D)$, $F$ is $B$, $G$ is $A \sqcap B$, $H$ is $B \vee (C \sqcup D)$, and $\gamma$ is '1.'.

**Proof.** Assume the conditions of the lemma. We proceed by induction on the surface complexity of $E$.

For the basis, assume $E$ is a quasiatom, so that $G = E$, $H = F$, and $\gamma$ is the empty string $\epsilon$. According to one of the assumptions of the lemma, $\langle\Phi_E^\epsilon\rangle G^* = F^*$. Hence, as $\Phi_E^\epsilon = \Phi$, we have $\langle\Phi\rangle G^* = F^*$. The equations $G = E$ and $H = F$ allow us to rewrite $\langle\Phi\rangle G^* = F^*$ as $\langle\Phi\rangle E^* = H^*$. Of course $H^* = \langle\rangle H^*$, and thus $\langle\Phi\rangle E^* = \langle\rangle H^*$. But notice that $\langle\rangle = \langle\Phi^{-\epsilon}\rangle$, which yields the desired $\langle\Phi\rangle E^* = \langle\Phi^{-\epsilon}\rangle H^*$.

Next, assume $E = \neg K$. As in the corresponding step of our proof of Lemma 3.12, $\gamma$ remains the $K$-specification of $G$, and $\Phi_E^\gamma = (\neg\Phi)_K^\gamma$. Also, $\neg\Phi \in \mathbf{LR}^{K^*}$. It is our assumption that $\langle\Phi_E^\gamma\rangle G^* = F^*$, and therefore $\langle(\neg\Phi)_K^\gamma\rangle G^* = F^*$. Then, by the induction hypothesis, $\langle\neg\Phi\rangle K^* = \langle(\neg\Phi)^{-\gamma}\rangle L^*$, where $L$ is the result of replacing $G$ by $F$ in $K$. Hence $\neg(\langle\neg\Phi\rangle K^*) = \neg(\langle(\neg\Phi)^{-\gamma}\rangle L^*)$. By Lemma 3.11(1), $\neg(\langle\neg\Phi\rangle K^*) = \langle\Phi\rangle E^*$ and $\neg(\langle(\neg\Phi)^{-\gamma}\rangle L^*) = \langle\Phi^{-\gamma}\rangle\neg L^*$. Consequently, $\langle\Phi\rangle E^* = \langle\Phi^{-\gamma}\rangle\neg L^*$. But, of course, $\neg L^* = H^*$. Thus, $\langle\Phi\rangle E^* = \langle\Phi^{-\gamma}\rangle H^*$.

Now, assume
$$E = K_1 \vee K_2 \vee \ldots \vee K_n.$$

Then, by Lemma 3.11(2),
$$\langle\Phi\rangle E^* = \langle\Phi^{1.}\rangle K_1^* \vee \langle\Phi^{2.}\rangle K_2^* \vee \ldots \vee \langle\Phi^{n.}\rangle K_n^*. \tag{3}$$

Let $K_i$ be the disjunct of $E$ that contains $G$, and let $\delta$ be the $K_i$-specification of $G$. For simplicity of representation and without loss of generality, let us assume here that $i = 1$. Thus, $\gamma = 1.\delta$, and we have

$$H = L \vee K_2 \vee \ldots \vee K_n, \tag{4}$$

where $L$ is the result of replacing $G$ by $F$ in $K_1$. As $\gamma = 1.\delta$, for any $j \neq 1$ we obviously have $\Phi^{j.} = (\Phi^{-\gamma})^{j.}$. Hence, (3) can be rewritten as

$$\langle\Phi\rangle E^* = \langle\Phi^{1.}\rangle K_1^* \vee \langle(\Phi^{-\gamma})^{2.}\rangle K_2^* \vee \ldots \vee \langle(\Phi^{-\gamma})^{n.}\rangle K_n^*. \tag{5}$$

It is our assumption that $\langle\Phi_E^\gamma\rangle G^* = F^*$, i.e. $\langle\Phi_E^{1.\delta}\rangle G^* = F^*$. But obviously $\Phi_E^{1.\delta} = (\Phi^{1.})_{K_1}^\delta$, and therefore $\langle(\Phi^{1.})_{K_1}^\delta\rangle G^* = F^*$. Then, by the induction hypothesis, $\langle\Phi^{1.}\rangle K_1^* = \langle(\Phi^{1.})^{-\delta}\rangle L^*$. But it is not hard to see that $(\Phi^{1.})^{-\delta} = (\Phi^{-1.\delta})^{1.}$. Hence, $\langle\Phi^{1.}\rangle K_1^* = \langle(\Phi^{-1.\delta})^{1.}\rangle L^*$, i.e. $\langle\Phi^{1.}\rangle K_1^* = \langle(\Phi^{-\gamma})^{1.}\rangle L^*$. This allows us to rewrite (5) as

$$\langle\Phi\rangle E^* = \langle(\Phi^{-\gamma})^{1.}\rangle L^* \vee \langle(\Phi^{-\gamma})^{2.}\rangle K_2^* \vee \ldots \vee \langle(\Phi^{-\gamma})^{n.}\rangle K_n^*. \tag{6}$$

Since $\Phi \in \mathbf{LR}^{E^*}$, every move of ($\Phi$ and hence of) $\Phi^{-\gamma}$ starts with '$i.$' for some $i \in \{1, \ldots, n\}$. And, with Convention 3.10 in mind, (6) implies that

$$(\Phi^{-\gamma})^{1.} \in \mathbf{LR}^{L^*}, \ (\Phi^{-\gamma})^{2.} \in \mathbf{LR}^{K_2^*}, \ \ldots, \ (\Phi^{-\gamma})^{n.} \in \mathbf{LR}^{K_n^*}.$$

By the definition of ∨, all this means that $\Phi^{-\gamma} \in \mathbf{LR}^{L^* \vee K_2^* \vee \ldots \vee K_n^*}$. Then, by Lemma 3.11(2),

$$\langle \Phi^{-\gamma} \rangle (L^* \vee K_2^* \vee \ldots \vee K_n^*) = \langle (\Phi^{-\gamma})^{1 \cdot} \rangle L^* \vee \langle (\Phi^{-\gamma})^{2 \cdot} \rangle K_2^* \vee \ldots \vee \langle (\Phi^{-\gamma})^{n \cdot} \rangle K_n^*.$$

The above, in conjunction with (6) and (4), yields the desired $\langle \Phi \rangle E^* = \langle \Phi^{-\gamma} \rangle H^*$.

Finally, in view of Lemma 3.11(3), the case $E = \exists x K$ is rather straightforward. This is so because $\mathbf{LR}^{\exists x K^*} = \mathbf{LR}^{K^*}$ and, as pointed out in our proof of Lemma 3.12, every quasiatom of $\exists x K$ is also a quasiatom of $K$ (and vice versa) with the same specification and same positive/negative status. □

## 3.4 Manageability

**Definition 3.14** Let $E$ be a reasonable hyperformula. We say that a run $\Gamma$ is $E$-**manageable** iff:

1. Every labmove of $\Gamma$ has the form $\wp\gamma\alpha$, where $\gamma$ $E$-specifies either a general or hybrid quasiatom.
2. Whenever $\pi$ is the $E$-specification of a positive hybrid quasiatom and $\nu$ is the $E$-specification of the same-base negative hybrid quasiatom, $\Gamma^\pi$ is a $\top$-delay of $\neg\Gamma^\nu$.
3. Whenever $\gamma$ is the $E$-specification of a general quasiatom, $\Gamma^\gamma$ contains no $\top$-labeled moves.

The above concept will play a central role in our soundness proof for **CL4**, with manageability being a certain nice property helping $\top$ to succeed. In rough intuitive terms, condition 1 means that the play has only been taking place in atoms, so that the logical structure of the game has not been affected, and it continues to be precisely described by $E$. Condition 2 ensures that the (sub)plays in the "matched" occurrences of atoms are in a sense symmetric, so that a win in at least one of them is guaranteed. And condition 3 signifies that $\top$ has not made any hasty moves in unmatched atoms, so that, if and when at some later point such an atom finds a match, by copying its adversary's moves, $\top$ will still have a chance to even out the corresponding two subplays.

**Lemma 3.15** *Assume $E$ is a reasonable hyperformula, $^*$ an $E$-admissible interpretation, and $\Gamma$ an infinite run with arbitrarily long finite initial segments that are $E$-manageable unilegal positions of $E^*$. Then $\Gamma$ is an $E$-manageable unilegal run of $E^*$.*

**Proof.** Assume the conditions of the lemma. By condition (a) of Definition 2.1 of [6], $\Gamma$ is in $\mathbf{LR}^{E^*}$ because it has arbitrarily long initial segments that are in $\mathbf{LR}^{E^*}$. And obviously $\Gamma$ satisfies conditions 1 and 3 of Definition 3.14 because it has arbitrarily long initial segments that satisfy those conditions. So, what remains to show is that $\Gamma$ also satisfies condition 2. Suppose, for a contradiction, that $\pi$ and $\nu$ are as in condition 2 but $\Gamma^\pi$ is not a $\top$-delay of $\neg\Gamma^\nu$. This means that at least one of the following two statements is true:

**(i)** For one of the players $\wp$, the subsequence of the $\wp$-labeled moves of $\Gamma^\pi$ (i.e. the result of deleting in $\Gamma^\pi$ all $\neg\wp$-labeled moves) is not the same as that of $\neg\Gamma^\nu$, or

**(ii)** For some $k, n$, in $\neg\Gamma^\nu$ the $n$th $\top$-labeled move is made later than the $k$th $\bot$-labeled move, but in $\Gamma^\pi$ the $n$th $\top$-labeled move is made earlier than the $k$th $\bot$-labeled move.

Whether (i) or (ii) is the case, it is not hard to see that, beginning from some (finite) $m$, every initial segment $\Psi$ of $\Gamma$ of length $\geq m$ will satisfy the same (i) or (ii) in the role of $\Gamma$, and hence $\Psi$ will not be an $E$-manageable position of $E^*$. This contradicts the assumptions of our lemma. □

**Lemma 3.16** *Assume $E$ is a reasonable hyperformula, $^*$ an $E$-admissible interpretation, and $\Omega$ an $E$-manageable unilegal position of $E^*$. Suppose $\gamma$ is the $E$-specification of a negative (resp. positive) quasiatom $G_1 \sqcap \ldots \sqcap G_n$ (resp. $G_1 \sqcup \ldots \sqcup G_n$), and $i \in \{1, \ldots, n\}$. Let $H$ be the result of replacing in $E$ the above quasiatom by $G_i$. Then:*

1. *$\Omega$ is $H$-manageable;*
2. *$\langle \Omega, \top\gamma i \rangle E^* = \langle \Omega \rangle H^*$.*

**Proof.** Assume the conditions of the lemma. It is not hard to see that, with $\Omega$ in the role of $\Gamma$, each of the thee conditions of Definition 3.14 is inherited by $H$ from $E$. This takes care of clause 1. Since $\Omega$ does not contain $\gamma$-prefixed moves (for, otherwise, by condition 1 of Definition 3.14, it would not be $E$-manageable), we have $\langle\Omega, \top\gamma i\rangle^\gamma = \langle\top i\rangle$ and $\langle\Omega, \top\gamma i\rangle^{-\gamma} = \Omega$. By clause 5 (resp. 6) of Lemma 4.7 of [6], the fact $\langle\Omega, \top\gamma i\rangle^\gamma = \langle\top i\rangle$ implies $\langle\Omega, \top\gamma i\rangle_E^\gamma (G_1 \sqcap \ldots \sqcap G_n)^* = G_i^*$ (resp. $\langle\Omega, \top\gamma i\rangle_E^\gamma (G_1 \sqcup \ldots \sqcup G_n)^* = G_i^*$); and Lemma 3.12 guarantees $\langle\Omega, \top\gamma i\rangle \in \mathbf{LR}^{E^*}$. Then, by Lemma 3.13, the fact $\langle\Omega, \top\gamma i\rangle^{-\gamma} = \Omega$ implies $\langle\Omega, \top\gamma i\rangle E^* = \langle\Omega\rangle H^*$. This proves clause 2. □

**Lemma 3.17** *Assume $E$ is a reasonable hyperformula, $*$ an $E$-admissible interpretation, and $\Omega$ an $E$-manageable unilegal position of $E^*$. Suppose $\gamma$ is the $E$-specification of a negative (resp. positive) quasiatom $\sqcap xG(x)$ (resp. $\sqcup xG(x)$), and $c$ is any constant. Let $H$ be the result of replacing in $E$ the above quasiatom by $G(c)$. Then:*
 1. *$\Omega$ is $H$-manageable;*
 2. *$\langle\Omega, \top\gamma c\rangle E^* = \langle\Omega\rangle H^*$.*

**Proof.** This lemma is very similar to the previous one. The only difference in the proof would be that, when claiming $\langle\Omega, \top\gamma c\rangle_E^\gamma \bigl(\sqcap xG(x)\bigr)^* = \bigl(G(c)\bigr)^*$ (resp. $\langle\Omega, \top\gamma c\rangle_E^\gamma \bigl(\sqcup xG(x)\bigr)^* = \bigl(G(c)\bigr)^*$), we would need to rely on clause 7 (resp. 8) of Lemma 4.7 of [6] in conjunction with Lemma 3.5. □

**Lemma 3.18** *Assume $A$ is a static game, $e$ is any valuation, and $\Gamma, \Delta$ are runs such that $\Delta$ is a $\top$-delay of $\Gamma$. Then:*
 1. *If $\Delta$ is a $\top$-illegal run of $e[A]$, then so is $\Gamma$.*
 2. *If $\Gamma$ is a $\bot$-illegal run of $e[A]$, then so is $\Delta$.*

**Proof.** The above is a fact known from [2] (Lemma 4.7). □

**Lemma 3.19** *Assume $E$ is a reasonable hyperformula, $*$ an $E$-admissible interpretation, and $\Omega$ an $E$-manageable unilegal position of $E^*$. Suppose $H$ is the hyperformula that results from $E$ by replacing in it a positive surface occurrence $\pi$ and a negative surface occurrence $\nu$ of a general letter $P$ by a hybrid letter $P_q$, such that $H$ remains reasonable. Further assume $\Omega^\pi = \langle\bot\pi_1, \ldots, \bot\pi_n\rangle$ and $\Omega^\nu = \langle\bot\nu_1, \ldots, \bot\nu_m\rangle$. Then $\langle\Omega, \top\pi\nu_1, \ldots, \top\pi\nu_m, \top\nu\pi_1, \ldots, \top\nu\pi_n\rangle$ is an $H$-manageable unilegal position of $H^*$.*

**Proof.** Assume the conditions of the lemma. Let
$$\Phi = \langle\Omega, \top\pi\nu_1, \ldots, \top\pi\nu_m, \top\nu\pi_1, \ldots, \top\nu\pi_n\rangle.$$

Notice that
$$\Phi_H^\pi = \Phi^\pi = \langle\bot\pi_1, \ldots, \bot\pi_n, \top\nu_1, \ldots, \top\nu_m\rangle \tag{7}$$
and
$$\Phi_H^\nu = \neg\Phi^\nu = \langle\top\nu_1, \ldots, \top\nu_m, \bot\pi_1, \ldots, \bot\pi_n\rangle. \tag{8}$$
As we see,
$$\Phi_H^\pi \text{ is a } \top\text{-delay of } \Phi_H^\nu. \tag{9}$$
This implies that $\Phi$ is $H$-manageable, because conditions 1 and 3 of Definition 3.14 are obviously inherited by $H$ and $\Phi$ from $E$ and $\Omega$, and so is condition 2 for any relevant pair $(\pi', \nu')$ different from $(\pi, \nu)$.

What remains to show is that $\Phi \in \mathbf{LR}^{H^*}$. For this, in view of Lemma 3.12, it would be sufficient to verify that $\Phi_H^\pi \in \mathbf{LR}^{P^*(\vec{t})}$ and $\Phi_H^\nu \in \mathbf{LR}^{P^*(\vec{t'})}$, where $P_q(\vec{t})$ and $P_q(\vec{t'})$ are the two — respectively positive and negative — $P_q$-based atoms of $E$. This is sufficient because for any other (different from $\pi, \nu$) relevant $\gamma$, the similar condition is inherited by $H$ and $\Phi$ from $E$ and $\Omega$ as we have $\Phi^\gamma = \Omega^\gamma$.

Since $\Omega \in \mathbf{LR}^{E^*}$, by Lemma 3.12 we have both $\Omega_E^\pi \in \mathbf{LR}^{P^*(\vec{t})}$ and $\Omega_E^\nu \in \mathbf{LR}^{P^*(\vec{t'})}$. Thus,
$$\langle\bot\pi_1, \ldots, \bot\pi_n\rangle \in \mathbf{LR}^{P^*(\vec{t})}; \tag{10}$$
$$\langle\top\nu_1, \ldots, \top\nu_m\rangle \in \mathbf{LR}^{P^*(\vec{t'})}. \tag{11}$$

Suppose $\Phi_H^\pi \notin \mathbf{LR}^{P^*(\vec{t})}$, i.e., for some valuation $e$, $\Phi_H^\pi \notin \mathbf{Lr}_e^{P^*(\vec{t})} = \mathbf{Lr}^{e[P^*(\vec{t})]}$. By (7), this can be rewritten as $\langle \bot\pi_1, \ldots, \bot\pi_n, \top\nu_1, \ldots, \top\nu_m \rangle \notin \mathbf{Lr}^{e[P^*(\vec{t})]}$. In view of (10), $\Phi_H^\pi$ cannot be a $\bot$-illegal position of $e[P^*(\vec{t})]$. So, it must be $\top$-illegal. But then, by (9) and Lemma 3.18(1), $\Phi_H^\nu$ is a $\top$-illegal position of $e[P^*(\vec{t})]$. Hence, by Lemma 3.4, $\Phi_H^\nu$ is a $\top$-illegal position of $e[P^*(\vec{t'})]$. This, however, is in obvious contradiction with (8) and (11).

Suppose now $\Phi_H^\nu \notin \mathbf{LR}^{P^*(\vec{t'})}$, i.e. $\Phi_H^\nu \notin \mathbf{Lr}^{e[P^*(\vec{t'})]}$ for some valuation $e$. This case is similar/symmetric to the previous one. By (8) and (11), $\Phi_H^\nu$ cannot be a $\top$-illegal position of $e[P^*(\vec{t'})]$. So, it must be $\bot$-illegal. But then, by (9) and Lemma 3.18(2), $\Phi_H^\pi$ is a $\bot$-illegal position of $e[P^*(\vec{t'})]$ and hence (by Lemma 3.4) of $e[P^*(\vec{t})]$. This is in contradiction with (7) and (10). $\square$

**Lemma 3.20** *Assume $E$ is a reasonable hyperformula, $\alpha$ is any move, $*$ is an $E$-admissible interpretation, $\Omega$ is an $E$-manageable position, and $\langle \Omega, \bot\alpha \rangle \in \mathbf{LR}^{E^*}$. Then one of the following conditions is satisfied:*

**(i)** *$\alpha = \gamma\beta$, where $\gamma$ is the $E$-specification of a general quasiatom. In this case $\langle \Omega, \bot\alpha \rangle$ is an $E$-manageable unilegal position of $E^*$.*

**(ii)** *$\alpha = \gamma\beta$, where $\gamma$ is the $E$-specification of a hybrid quasiatom. Let $\sigma$ be the $E$-specification of the other same-base hybrid quasiatom. Then $\langle \Omega, \bot\gamma\beta, \top\sigma\beta \rangle$ is an $E$-manageable unilegal position of $E^*$.*

**(iii)** *$\alpha = \gamma i$, where $\gamma$ is the $E$-specification of a positive (resp. negative) quasiatom $G_1 \sqcap \ldots \sqcap G_n$ (resp. $G_1 \sqcup \ldots \sqcup G_n$) and $i \in \{1, \ldots, n\}$. In this case, where $H$ is the result of replacing in $E$ the above quasiatom by $G_i$, we have:*
   1. *$\Omega$ is $H$-manageable;*
   2. *$\langle \Omega, \bot\alpha \rangle E^* = \langle \Omega \rangle H^*$.*

**(iv)** *$\alpha = \gamma c$, where $\gamma$ is the $E$-specification of a positive (resp. negative) quasiatom $\sqcap xG(x)$ (resp. $\sqcup xG(x)$) and $c$ is a constant. In this case, where $H$ is the result of replacing in $E$ the above quasiatom by $G(c)$, we have:*
   1. *$\Omega$ is $H$-manageable;*
   2. *$\langle \Omega, \bot\alpha \rangle E^* = \langle \Omega \rangle H^*$.*

**Proof.** Assume $E$, $\alpha$, $*$, $\Omega$ are as the first sentence of the lemma stipulates. By Lemma 3.12, the condition $\langle \Omega, \bot\alpha \rangle \in \mathbf{LR}^{E^*}$ implies that $\alpha$ should be $\gamma\beta$, where $\gamma$ is a $E$-specification of a non-elementary quasiatom $F$ of $E$. Fix these $\gamma, \beta$ and $F$. As a non-elementary quasiatom, $F$ should be either (i) a general atom, or (ii) a hybrid atom, or (iii) a $\sqcap$- or $\sqcup$-(hyper)formula, or (iv) a $\sqcap$- or $\sqcup$-(hyper)formula. We consider each of these four possibilities — corresponding to the four conditions (i)-(iv) of the lemma — separately.

*Case (i):* $F$ is a general atom. Obviously adding to a $E$-manageable position ($\Omega$) a $\bot$-labeled move whose prefix $E$-specifies a general quasiatom again yields an $E$-manageable position. So, $\langle \Omega, \bot\alpha \rangle$ is $E$-manageable; by the assumptions of the lemma, it is also a unilegal position of $E^*$. Thus, condition (i) of Lemma 3.20 is satisfied.

*Case (ii):* $F$ is a hybrid atom $P_q(\vec{t})$. Let $P_q(\vec{t'})$ be the other $P_q$-based quasiatom of $E$ and $\sigma$ its $E$-specification, so that $\gamma, \sigma$ are as in (the condition of) condition (ii) of the lemma. We want to show that then the rest of that condition is also satisfied, i.e. that $\langle \Omega, \bot\gamma\beta, \top\sigma\beta \rangle$ is an $E$-manageable unilegal position of $E^*$.

*Subcase (ii).1:* Assume $P_q(\vec{t})$ is negative in $E$ and $P_q(\vec{t'})$ is positive. Since $\Omega$ is $E$-manageable, $\Omega^\sigma$ is a $\top$-delay of $\neg\Omega^\gamma$. Therefore $\langle \Omega^\sigma, \top\beta \rangle$ is a $\top$-delay of $\langle \neg\Omega^\gamma, \top\beta \rangle$, i.e. of $\neg\langle \Omega^\gamma, \bot\beta \rangle$. From here, observing that $\langle \Omega^\sigma, \top\beta \rangle = \langle \Omega, \bot\gamma\beta, \top\sigma\beta \rangle^\sigma$ and $\langle \Omega^\gamma, \bot\beta \rangle = \langle \Omega, \bot\gamma\beta, \top\sigma\beta \rangle^\gamma$, we get:

$$\langle \Omega, \bot\gamma\beta, \top\sigma\beta \rangle^\sigma \text{ is a } \top\text{-delay of } \neg\langle \Omega, \bot\gamma\beta, \top\sigma\beta \rangle^\gamma. \tag{12}$$

Remembering the assumption that $\Omega$ is $E$-manageable and taking into account that for any relevant $\delta \neq \gamma, \sigma$ we have $\langle \Omega, \bot\gamma\beta, \top\sigma\beta \rangle^\delta = \Omega^\delta$, (12) is obviously sufficient to conclude that

$$\langle \Omega, \bot\gamma\beta, \top\sigma\beta \rangle \text{ is } E\text{-manageable.} \tag{13}$$

According to the assumptions of the lemma, $\langle \Omega, \bot\gamma\beta \rangle \in \mathbf{LR}^{E^*}$. By Lemma 3.12, this implies that $\langle \Omega, \bot\gamma\beta \rangle_E^\gamma \in \mathbf{LR}^{P^*(\vec{t})}$, i.e. $\neg\langle \Omega, \bot\gamma\beta \rangle^\gamma \in \mathbf{LR}^{P^*(\vec{t})}$. But $\neg\langle \Omega, \bot\gamma\beta \rangle^\gamma = \neg\langle \Omega, \bot\gamma\beta, \top\sigma\beta \rangle^\gamma$ because $\gamma \neq \sigma$.

Thus, $\neg\langle\Omega, \bot\gamma\beta, \top\sigma\beta\rangle^\gamma \in \mathbf{LR}^{P^*(\vec{t})}$. Then, by (12) and Lemma 3.18(1), $\langle\Omega, \bot\gamma\beta, \top\sigma\beta\rangle^\sigma$ is not a $\top$-illegal position of $e[P^*(\vec{t})]$ (whatever valuation $e$) and hence — by Lemma 3.4 — of $e[P^*(\vec{t'})]$. It is not a $\bot$-illegal position of $e[P^*(\vec{t'})]$ either, for otherwise we would have $\langle\Omega, \bot\gamma\beta\rangle^\sigma \notin \mathbf{LR}^{P^*(\vec{t'})}$ which, in view of Lemma 3.12, contradicts our assumption that $\langle\Omega, \bot\gamma\beta\rangle \in \mathbf{LR}^{E^*}$. Thus, $\langle\Omega, \bot\gamma\beta, \top\sigma\beta\rangle^\sigma \in \mathbf{LR}^{P^*(\vec{t'})}$. Now, taking into account that for any $\delta \neq \sigma$ that $E$-specifies a quasiatom of $E$ we clearly have $\langle\Omega, \bot\gamma\beta, \top\sigma\beta\rangle^\delta = \langle\Omega, \bot\gamma\beta\rangle^\delta$, Lemma 3.12 in conjunction with our assumption $\langle\Omega, \bot\gamma\beta\rangle \in \mathbf{LR}^{E^*}$ implies that $\langle\Omega, \bot\gamma\beta, \top\sigma\beta\rangle \in \mathbf{LR}^{E^*}$. This, together with (13), means that condition (ii) of Lemma 3.20 is satisfied.

*Subcase (ii).2:* Now assume $P_q(\vec{t})$ is positive in $E$ and $P_q(\vec{t'})$ is negative. By the $E$-manageability of $\Omega$, $\Omega^\gamma$ is a $\top$-delay of $\neg\Omega^\sigma$, whence $\langle\Omega^\gamma, \bot\beta\rangle$ is a $\top$-delay of $\langle\neg\Omega^\sigma, \bot\beta\rangle = \neg\langle\Omega^\sigma, \top\beta\rangle$. Then, as $\langle\Omega^\gamma, \bot\beta\rangle = \langle\Omega, \bot\gamma\beta, \top\sigma\beta\rangle^\gamma$ and $\langle\Omega^\sigma, \top\beta\rangle = \langle\Omega, \bot\gamma\beta, \top\sigma\beta\rangle^\sigma$, we get:

$$\langle\Omega, \bot\gamma\beta, \top\sigma\beta\rangle^\gamma \text{ is a } \top\text{-delay of } \neg\langle\Omega, \bot\gamma\beta, \top\sigma\beta\rangle^\sigma. \tag{14}$$

From here, just as from (12) in Subcase (ii).1, we can conclude that statement (13) is true.

Next, as noted in Subcase (ii).1, $\langle\Omega, \bot\gamma\beta\rangle^\gamma_E \in \mathbf{LR}^{P^*(\vec{t})}$, which now simply means that $\langle\Omega, \bot\gamma\beta\rangle^\gamma \in \mathbf{LR}^{P^*(\vec{t})}$. Then $\langle\Omega, \bot\gamma\beta, \top\sigma\beta\rangle^\gamma$, which equals $\langle\Omega, \bot\gamma\beta\rangle^\gamma$, is not a $\bot$-illegal position of $e[P^*(\vec{t})]$ (whatever valuation $e$). Therefore, by (14) and Lemma 3.18(2), $\neg\langle\Omega, \bot\gamma\beta, \top\sigma\beta\rangle^\sigma$ is not a $\bot$-illegal position of $e[P^*(\vec{t})]$ and hence, by Lemma 3.4, of $e[P^*(\vec{t'})]$. But $\neg\langle\Omega, \bot\gamma\beta, \top\sigma\beta\rangle^\sigma$, being equal to $\langle\neg\Omega, \top\gamma\beta, \bot\sigma\beta\rangle^\sigma$, is not a $\top$-illegal position of $e[P^*(\vec{t'})]$, either, for otherwise we would have $\langle\neg\Omega, \top\gamma\beta\rangle^\sigma \notin \mathbf{LR}^{P^*(\vec{t'})}$, i.e. $\neg\langle\Omega, \bot\gamma\beta\rangle^\sigma \notin \mathbf{LR}^{P^*(\vec{t'})}$, i.e. $\langle\Omega, \bot\gamma\beta\rangle^\sigma_E \notin \mathbf{LR}^{P^*(\vec{t'})}$, and this, by Lemma 3.12, contradicts our assumption that $\langle\Omega, \bot\gamma\beta\rangle \in \mathbf{LR}^{E^*}$. Thus, $\langle\Omega, \bot\gamma\beta, \top\sigma\beta\rangle^\sigma \in \mathbf{LR}^{P^*}(\vec{t'})$. From here, as in Subcase (ii).1, we can conclude that condition (ii) of Lemma 3.20 is satisfied.

*Case (iii):* $F$ is $G_1 \sqcap \ldots \sqcap G_n$ or $G_1 \sqcup \ldots \sqcup G_n$. We want to show that then (the rest of) condition (iii) of Lemma 3.20 is satisfied. Since $\Omega$ is $E$-manageable, clause 1 of Definition 3.14 implies that $\Omega$ does not contain $\gamma$-prefixed moves. Therefore we have:

$$\langle\Omega, \bot\gamma\beta\rangle^\gamma = \langle\bot\gamma\beta\rangle^\gamma; \tag{15}$$
$$\langle\Omega, \bot\gamma\beta\rangle^{\neg\gamma} = \Omega. \tag{16}$$

By (15) and Lemma 3.12, $\langle\bot\gamma\beta\rangle^\gamma_E \in \mathbf{LR}^{F^*}$. In view of clauses 5(a) and 6(a) of Lemma 4.7 of [6], this is the case when $\beta = i \in \{1, \ldots, n\}$ and either $F = G_1 \sqcap \ldots \sqcap G_n$ and $F$ is positive in $E$ (so that $\langle\bot\gamma\beta\rangle^\gamma_E = \langle\bot i\rangle$), or $F = G_1 \sqcup \ldots \sqcup G_n$ and $F$ is negative in $E$ (so that $\langle\bot\gamma\beta\rangle^\gamma_E = \langle\top i\rangle$). In either case, by clauses 5(b) and 6(b) of the same lemma, we have $\langle\bot\gamma\beta\rangle^\gamma_E F^* = G_i^*$. By (15), this means that $\langle\Omega, \bot\gamma\beta\rangle^\gamma_E F^* = G_i^*$. Then, according to Lemma 3.13, $\langle\Omega, \bot\gamma\beta\rangle E^* = \langle\Omega, \bot\gamma\beta\rangle^{\neg\gamma} H^*$, where $H$ is the result of replacing in $E$ the quasiatom $F$ by $G_i$. Applying (16) and changing $\gamma\beta$ back to $\alpha$, the just-derived equation can be rewritten as $\langle\Omega, \bot\alpha\rangle E^* = \langle\Omega\rangle H^*$. To conclude that condition (iii) of Lemma 3.20 is satisfied, it remains to notice that $\Omega$ is $H$-manageable. This is so for the same reasons as in Lemma 3.16.

*Case (iv):* $F$ is $\sqcap xG(x)$ or $\sqcup xG(x)$. This case is similar to the previous one, only we need to use clauses 7 and 8 (instead of 5 and 6) of Lemma 4.7 of [6], and also rely on Lemma 3.5 when doing so. □

## 3.5 Finalization

We define the relation $\leq$ on elementary games by stipulating that $A \leq B$ iff, for every valuation $e$, whenever $e[A]$ is true, so is $e[B]$. Informally this can be read as "$B$ is at least as true as $A$".

**Lemma 3.21** *Assume $p_1$ and $p_2$ are same-arity elementary letters; $F_1$ is an elementary formula; $F_2$ is the result of replacing in $F_1$ a positive (resp. negative) occurrence of $p_1$ by $p_2$; and $^*$ is an interpretation such that $p_1^*$ and $p_2^*$ have the same canonical tuple, with $p_2^* \leq p_1^*$ (resp. $p_1^* \geq p_2^*$). Then $F_2^* \leq F_1^*$.*

**Idea.** The above lemma, in fact, does nothing but rephrases, in our terms, a well known fact from classical logic. Let us not be lazy to still verify it for safety.

**Proof.** Assume the conditions of the lemma. Let $(x_1, \ldots, x_n)$ be the canonical tuple of $p_1^*$ and $p_2^*$, and $p_1(t_1, \ldots, t_n)$ be the atom of $F_1$ in which $p_1$ got replaced by $p_2$ when obtaining $F_2$. We proceed by induction on the complexity of $F_1$. As was done earlier, the inductive step will be limied to the cases when the main operator of $F_1$ is $\neg$, $\vee$ or $\exists$.

If $F_1$ is atomic, then $F_1 = p_1(t_1, \ldots, t_n)$ and $F_2 = p_2(t_1, \ldots, t_n)$. The occurrence of $p_1$ in $F_1$ is thus positive and hence, by the conditions of the lemma, $p_2^* \leq p_1^*$. Pick any valuation $e$. Let $g$ be the valuation that agrees with $e$ on all variables that are not among $x_1, \ldots, x_n$, such that $g(x_1) = e(t_1), \ldots, g(x_n) = e(t_n)$. By Definition 4.1 of [6], $e[p_1^*(t_1, \ldots, t_n)] = g[p_1^*(x_1, \ldots, x_n)]$ and $e[p_2^*(t_1, \ldots, t_n)] = g[p_2^*(x_1, \ldots, x_n)]$. But $p_2^* \leq p_1^*$ implies that, whenever $g[p_2^*(x_1, \ldots, x_n)]$ is true, so is $g[p_1^*(x_1, \ldots, x_n)]$; hence, whenever $e[p_2^*(t_1, \ldots, t_n)]$ is true, so is $e[p_1^*(t_1, \ldots, t_n)]$. As $e$ was arbitrary, we conclude that $F_2^* \leq F_1^*$.

Suppose now $F_1 = \neg G_1$. Then $F_2 = \neg G_2$, where the conditions of the lemma are satisfied with $G_2, G_1, p_2, p_1$ in the role of $F_1, F_2, p_1, p_2$, respectively (the subscripts $_1$ and $_2$ have been interchanged because what was positive in $F_1, F_2$, is negative in $G_1, G_2$, and vice versa). Hence, by the induction hypothesis, $G_1^* \leq G_2^*$. This obviously means $\neg G_2^* \leq \neg G_1^*$, i.e. $F_2^* \leq F_1^*$.

Next, suppose $F_1$ is a disjunction. Without loss of generality, we may assume that the replaced occurrence of $p_1$ happens to be in the first disjunct. So, $F_1$ has the form $G_1 \vee H_1 \vee \ldots \vee H_n$ and $F_2$ is $G_2 \vee H_1 \vee \ldots \vee H_n$, where the conditions of the lemma are satisfied with $G_1$ and $G_2$ in the role of $F_1$ and $F_2$, respectively. By the induction hypothesis, $G_2^* \leq G_1^*$. Obviously this implies $F_2^* \leq F_1^*$.

Finally, suppose $F_1 = \exists x G_1$, so that we have $F_2 = \exists x G_2$, where the conditions of the lemma are satisfied with $G_1$ and $G_2$ in the role of $F_1$ and $F_2$, respectively. Pick any valuation $e$ and assume $e[\exists x G_2^*]$ is true. By the definition of $\exists$, this means that $g[G_2^*]$ is true for some valuation $g$ that agrees with $e$ on all variables except perhaps $x$. But, by the induction hypothesis, $G_2^* \leq G_1^*$, so that, for that $g$, $g[G_1^*]$ is also true. Consequently, $e[\exists x G_1^*]$ is true. Since $e$ was arbitrary, we conclude that $F_2^* \leq F_1^*$. □

Now we introduce the operation $\langle \Gamma \rangle \downarrow A$ of the type $\{runs\} \times \{games\} \to \{elemntary\ games\}$, which is rather similar to prefixation. The meaning of the predicate $\langle \Gamma \rangle \downarrow A$, which we call the $\Gamma$-**finalization** of $A$, is "$\Gamma$ is a $\top$-won run of $e[A]$". That is, $\langle \Gamma \rangle \downarrow A$ is true at $e$ iff $\mathbf{Wn}_e^A \langle \Gamma \rangle = \top$. This operation is only defined when $\Gamma$ is a unilegal run of $A$. Here is an official definition:

**Definition 3.22** Let $A$ be any game and $\Gamma$ a unilegal run of $A$. $\langle \Gamma \rangle \downarrow A$ is defined by: $\mathbf{Lr}_e^{\langle \Gamma \rangle \downarrow A} = \{\langle\rangle\}$; $\mathbf{Wn}_e^{\langle \Gamma \rangle \downarrow A} \langle\rangle = \mathbf{Wn}_e^A \langle \Gamma \rangle$.

Extending Convention 3.10 to finalization, every time we write $\langle \Gamma \rangle \downarrow A$, we implicitly claim that this game is defined, i.e. $\Gamma \in \mathbf{LR}^A$. The follwoing fact is immediate from the relevant definitions:

**Lemma 3.23** For any games $A, A_1, \ldots, A_n$ $(n \geq 2)$ we have:
  1. $\langle\rangle \downarrow (A_1 \sqcap \ldots \sqcap A_n) = \langle\rangle \downarrow \sqcap x A = \top$.
  2. $\langle\rangle \downarrow (A_1 \sqcup \ldots \sqcup A_n) = \langle\rangle \downarrow \sqcup x A = \bot$.

The following lemma, which is very similar to (but somewhat simpler than) Lemma 3.11, can be verified by a routine analysis of the relevant definitions:

**Lemma 3.24** Let $A, A_1, \ldots, A_n$ $(n \geq 2)$ be any games, $x$ any variable, and $\Gamma$ any run. For each of the following clauses, we assume that $\Gamma$ is a unilegal run of the game to which $\Gamma$-finalization is applied; for clause 3 we also assume that $A$ is $x$-unistructural. Then:
  1. $\langle \Gamma \rangle \downarrow \neg A = \neg(\langle \neg \Gamma \rangle \downarrow A)$.
  2. $\langle \Gamma \rangle \downarrow (A_1 \vee \ldots \vee A_n) = \langle \Gamma^{1.} \rangle \downarrow A_1 \vee \ldots \vee \langle \Gamma^{n.} \rangle \downarrow A_n$.
  3. $\langle \Gamma \rangle \downarrow \exists x A = \exists x \langle \Gamma \rangle \downarrow A$.

Immediately based on Definition 3.22 and the definition of substitution of variables (Definition 4.1 of [6]), we also easily find that the following is true:

**Lemma 3.25** Let $A$ be a game, $\vec{x}$ an n-tuple of pairwise distinct variables, $\vec{t}$ any n-tuple of terms, and $\Gamma$ a unilegal run of $A$. Then $(\langle \Gamma \rangle \downarrow A)[\vec{x}/\vec{t}] = \langle \Gamma \rangle \downarrow (A[\vec{x}/\vec{t}])$.

**Lemma 3.26** Assume $E, F$ are hyperformulas, $G$ is a quasiatom of $E$, and $H$ is the result of replacing this quasiatom by $F$ in $E$. Further assume $\gamma$ is the $E$-specification of $G$, $^*$ is an interpretation admissible for both $E$ and $H$, and $\Gamma$ is a unilegal run of $E^*$ with $\langle \Gamma_E^\gamma \rangle \downarrow G^* = F^*$. Then $\langle \Gamma \rangle \downarrow E^* = \langle \Gamma^{\neg \gamma} \rangle \downarrow H^*$.

**Proof.** This lemma is very similar to Lemma 3.13, and the proof of the latter can be literally repeated here as long as in it we replace $\Phi$ by $\Gamma$, prefixation by finalization, and references to Lemma 3.11 by references to Lemma 3.24. $\square$

**Lemma 3.27** *Assume $E$ is a stable, reasonable, closed hyperformula, $*$ is an $E$-admissible perfect interpretation,[5] and $\Gamma$ is an $E$-manageable (uni)legal run of $E^*$. Then $\mathbf{Wn}^{E^*}\langle\Gamma\rangle = \top$.*

**Idea.** In the role of $E$, let us take the obviously stable hyperformula $S \vee \neg P_q \vee ((P_q \wedge \sqcap xQ(x) \wedge (r \vee \neg r))$. Consider the $E$-manageable run $\Gamma = \langle \bot 1.\alpha, \bot 2.\beta, \bot 3.1.\delta, \top 2.\delta, \top 3.1.\beta\rangle$, and assume it is in $\mathbf{LR}^E$ (by abuse of concepts, here we identify hyperformulas with their interpretations). Applying Lemma 3.26 as many times as the number of non-elementary quasiatoms of $E$, we find that $\langle\Gamma\rangle\downarrow E$ is the proposition

$$(\langle\bot\alpha\rangle\downarrow S) \vee \neg(\langle\top\beta, \bot\delta\rangle\downarrow P_q) \vee \Big(\big((\langle\bot\delta, \top\beta\rangle\downarrow P_q)\big) \wedge (\langle\rangle\downarrow\sqcap xQ(x)) \wedge (r \vee \neg r)\Big). \tag{17}$$

Remembering the meaning of finalization, our lemma's claim that $\Gamma$ is a $\top$-won run of $E$ means nothing but that $\langle\Gamma\rangle\downarrow E$, i.e. (17), is true. So, what we want to understand is why (17) is guaranteed to be true. Let us look at the elementarization

$$\bot \vee \neg q \vee \big(q \wedge \top \wedge (r \vee \neg r)\big) \tag{18}$$

of $E$. It is a tautology due to the stability of $E$. We now want to see that (17) has to be "at least as true" as (18). First of all, note that, by Lemma 3.23, the $\langle\rangle\downarrow\sqcap xQ(x)$ part of (17) is identical to the corresponding part $\top$ of (18). The applicability of Lemma 3.23 here is not just a lucky coincidence, and in the general case is guaranteed by clause 1 of Definition 3.14, according to which a manageable run would never contain moves made in choice subcomponents of the game such as — in our case — $\sqcap xQ(x)$. Next, while the $\langle\bot\alpha\rangle\downarrow S$ part of (17) is not necessarily the same as the corresponding part $\bot$ of (18), at least the former can never be "more false" than the latter, so it is fine if we show the truth of (17) while generously seeing in it $\bot$ instead of $\langle\bot\alpha\rangle\downarrow S$. Again, remembering what elementarization does to general atoms such as — in our case — $S$, this sort of a "not more false" situation with general-letter-related components is not an accident. Thus, what would suffice now to see is the truth of

$$\bot \vee \neg\big(\langle\top\beta, \bot\delta\rangle\downarrow P_q\big) \vee \Big(\big((\langle\bot\delta, \top\beta\rangle\downarrow P_q)\big) \wedge \top \wedge (r \vee \neg r)\Big). \tag{19}$$

Let us look at the following formula, which is a substitutional instance of the tautological (18) and hence true:

$$\bot \vee \neg\big(\langle\top\beta, \bot\delta\rangle\downarrow P_q\big) \vee \Big(\big((\langle\top\beta, \bot\delta\rangle\downarrow P_q)\big) \wedge \top \wedge (r \vee \neg r)\Big). \tag{20}$$

The two formulas (19) and (20) only differ in the first conjunct of the third disjunct. So, to conclude that (19) is true, it would be sufficient to see that $\langle\bot\delta, \top\beta\rangle \downarrow P_q$ is "at least as true" as $\langle\top\beta, \bot\delta\rangle \downarrow P_q$. But this is indeed so: by condition 2 of Definition 3.14, $\langle\bot\delta, \top\beta\rangle$ is a $\top$-delay of $\langle\top\beta, \bot\delta\rangle$, and therefore — by the property of static games — the former is won by $\top$ as long as the latter is so; in other words, $\langle\top\beta, \bot\delta\rangle\downarrow P_q \leq \langle\bot\delta, \top\beta\rangle\downarrow P_q$.

**Proof.** Assume the conditions of the lemma. Since $\Gamma \in \mathbf{LR}^{E^*}$, by Lemma 3.12, every labmove of $\Gamma$ has the form $\wp\gamma\beta$, where $\gamma$ is the $E$-specification of a non-elementary quasiatom $G$. If such a $G$ is $H_1 \sqcap \ldots \sqcap H_n$ or $\sqcap xH$, as $\Gamma$ is $E$-manageable, we have $\Gamma_E^\gamma = \langle\rangle$ and hence, by clause 1 of Lemma 3.23, $\langle\Gamma_E^\gamma\rangle\downarrow G^* = \top$. Similarly, if $G$ is $H_1 \sqcup \ldots \sqcup H_n$ or $\sqcup xH$, clause 2 of the same lemma yields $\langle\Gamma_E^\gamma\rangle\downarrow G^* = \bot$. Thus we have:

$$\begin{array}{l} \textit{If } \gamma \textit{ E-specifies a } \sqcap\textit{- or } \sqcap\textit{-quasiatom } G, \textit{ then } \langle\Gamma_E^\gamma\rangle\downarrow G^* = \top^*. \\ \textit{If } \gamma \textit{ E-specifies a } \sqcup\textit{- or } \sqcup\textit{-quasiatom } G, \textit{ then } \langle\Gamma_E^\gamma\rangle\downarrow G^* = \bot^*. \end{array} \tag{21}$$

For each string $\gamma$ that $E$-specifies a general or hybrid quasiatom $G$, let us fix a non-logical elementary letter $r_\gamma$ not occurring in $E$ (neither directly nor as the elementary component of a hybrid letter). We require that the arity of such an $r_\gamma$ be 0 if $G$ is a general atom, and be the same as the arity of $G$ if the latter happens to be a hybrid atom. We also require that all these $r_\gamma$ be pairwise distinct. The official definition views interpretations as total functions — ones defined for all letters of the language. This is only for convenience, and obviously nothing will change if we assume in this proof that originally $*$ is only defined for the letters occurring in $E$. Then we can extend $*$ to other letters — specifically, the above $r_\gamma$ — in an arbitrary way that serves our goal. This is what will be done in the following two paragraphs.

---

[5]In fact, the conditions that $E$ is closed and $*$ is perfect could be dropped here, but why bother.

For a $\gamma$ that $E$-specifies a general quasiatom $G$, we let $r_\gamma^*$ be $\langle\Gamma_E^\gamma\rangle \downarrow G^*$. The latter is defined, i.e. $\Gamma_E^\gamma \in \mathbf{LR}^{G^*}$, by Lemma 3.12 and our assumption that $\Gamma \in \mathbf{LR}^{E^*}$. Let us mark this for future references:

$$\text{If } \gamma \text{ } E\text{-specifies a general quasiatom } G, \text{ then } \langle\Gamma_E^\gamma\rangle \downarrow G^* = r_\gamma^*. \tag{22}$$

Next, consider a $\gamma$ that $E$-specifies an $n$-ary hybrid quasiatom $P_q(\vec{t})$, with $\vec{t}$ abbreviating $t_1, \ldots, t_n$. Let $P_q(\vec{t'}) = P_q(t_1', \ldots, t_n')$ be the other $P_q$-based quasiatom of $E$. Assume $(\vec{x}) = (x_1, \ldots, x_n)$ is the canonical tuple of $P^*$. Let $\vec{y}$ be the result of deleting in $x_1, \ldots, x_n$ every $x_i$ such that at least one of the terms $t_i, t_i'$ is a variable. And let $\vec{z}$ be the result of deleting in $x_1, \ldots, x_n$ every $x_i$ such that both $t_i$ and $t_i'$ are constants. For convenience of representation and without loss of generality, we assume here that all variables from $\vec{y}$ go before all variables from $\vec{z}$ in $\vec{x}$; that is, for some $m$ (fix it) with $1 \leq m \leq n$, we have $\vec{y} = x_1, \ldots, x_m$ and $\vec{z} = x_{m+1}, \ldots, x_n$. So, we can rewrite $P^*(\vec{x})$ as $P^*(\vec{y}, \vec{z})$. The reasonabity of $E$ implies that $t_{m+1} = t_{m+1}'$, $\ldots, t_n = t_n'$. Let $\vec{c} = t_{m+1}, \ldots, t_n$, $\vec{h} = t_1, \ldots, t_m$ and $\vec{h}' = t_1', \ldots, t_m'$. So, we can rewrite $P_q(\vec{t})$ as $P_q(\vec{h}, \vec{c})$ and $P_q(\vec{t'})$ as $P_q(\vec{h}', \vec{c})$. As we agreed, for each $i \in \{1, \ldots, m\}$, either $t_i$ or $t_i'$ is a variable and hence — since $E$ is closed — its occurrence within $P_q(\vec{t})$ or $P_q(\vec{t'})$ is blindly bound (it cannot be bound by a choice quantifier because the occurrences of $P_q(\vec{t})$ and $P_q(\vec{t'})$ are surface ones). Then the $E$-admissibility of $*$ implies that $P^*(\vec{y}, \vec{z})$ — and hence obviously $P^*(\vec{y}, \vec{c})$ — is unistructural in each of the variables of $\vec{y}$. Based on this, it is not hard to see that the games $P^*(\vec{y}, \vec{c})$ and $P^*(\vec{h}, \vec{c})$ are equistructural. But, since $\Gamma \in \mathbf{LR}^{E^*}$, by Lemma 3.12, we have $\Gamma_E^\gamma \in \mathbf{LR}^{P^*(\vec{h}, \vec{c})}$. Consequently, $\Gamma_E^\gamma \in \mathbf{LR}^{P^*(\vec{y}, \vec{c})}$, and hence the game $\langle\Gamma_E^\gamma\rangle \downarrow P^*(\vec{y}, \vec{c})$ is defined. This makes it possible for us to extend $*$ to $r_\gamma$ by stipulating that $*$ interprets it as the predicate $r_\gamma^*(\vec{y}, \vec{z})$, with

$$r_\gamma^*(\vec{y}, \vec{z}) = \langle\Gamma_E^\gamma\rangle \downarrow P^*(\vec{y}, \vec{c}).$$

Note that $r_\gamma^*(\vec{y}, \vec{z})$ does not depend on $\vec{z}$ which is just a "place filler" to match the requirement that $r_\gamma$ is $n$-ary.[6] Now:

- $\langle\Gamma_E^\gamma\rangle \downarrow \big(P_q(\vec{t})\big)^* = \langle\Gamma_E^\gamma\rangle \downarrow P^*(\vec{h}, \vec{c})$ — just rewriting $\big(P_q(\vec{t})\big)^*$ as $P^*(\vec{t})$ and then $(\vec{t})$ as $(\vec{h}, \vec{c})$;

- $\langle\Gamma_E^\gamma\rangle \downarrow P^*(\vec{h}, \vec{c}) = \langle\Gamma_E^\gamma\rangle \downarrow \Big(\big(P^*(\vec{y}, \vec{c})\big)[\vec{y}/\vec{h}]\Big)$ — rewriting $P^*(\vec{h}, \vec{c})$ as $\big(P^*(\vec{y}, \vec{c})\big)[\vec{y}/\vec{h}]$;

- $\langle\Gamma_E^\gamma\rangle \downarrow \Big(\big(P^*(\vec{y}, \vec{c})\big)[\vec{y}/\vec{h}]\Big) = \langle\Gamma_E^\gamma\rangle \downarrow \Big(\big(P^*(\vec{y}, \vec{c})\big)[\vec{y}/\vec{h}, \vec{z}/\vec{c}]\Big)$ — vacuously as $P^*(\vec{y}, \vec{c})$ does not depend on $\vec{z}$ and hence $\big(P^*(\vec{y}, \vec{c})\big)[\vec{y}/\vec{h}] = \big(P^*(\vec{y}, \vec{c})\big)[\vec{y}/\vec{h}, \vec{z}/\vec{c}]$;

- $\langle\Gamma_E^\gamma\rangle \downarrow \Big(\big(P^*(\vec{y}, \vec{c})\big)[\vec{y}/\vec{h}, \vec{z}/\vec{c}]\Big) = \Big(\langle\Gamma_E^\gamma\rangle \downarrow \big(P^*(\vec{y}, \vec{c})\big)\Big)[\vec{y}/\vec{h}, \vec{z}/\vec{c}]$ — by Lemma 3.25;

- $\Big(\langle\Gamma_E^\gamma\rangle \downarrow \big(P^*(\vec{y}, \vec{c})\big)\Big)[\vec{y}/\vec{h}, \vec{z}/\vec{c}] = \big(r_\gamma^*(\vec{y}, \vec{z})\big)[\vec{y}/\vec{h}, \vec{z}/\vec{c}]$ — by our definition of $r_\gamma^*$;

- $\big(r_\gamma^*(\vec{y}, \vec{z})\big)[\vec{y}/\vec{h}, \vec{z}/\vec{c}] = r_\gamma^*(\vec{t})$ — just rewriting $\big(r_\gamma^*(\vec{y}, \vec{z})\big)[\vec{y}/\vec{h}, \vec{z}/\vec{c}]$ as $r_\gamma^*(\vec{h}, \vec{c})$ and then $(\vec{h}, \vec{c})$ as $(\vec{t})$.

The above chain of equations yields $\langle\Gamma_E^\gamma\rangle \downarrow \big(P_q(\vec{t})\big)^* = r_\gamma^*(\vec{t})$. Generalizing to all $\gamma$, we thus have:

$$\text{If } \gamma \text{ } E\text{-specifies a hybrid quasiatom } P_q(\vec{t}), \text{ then } \langle\Gamma_E^\gamma\rangle \downarrow \big(P_q(\vec{t})\big)^* = r_\gamma^*(\vec{t}). \tag{23}$$

**Claim 1** *Assume $\pi$ is the $E$-specification of a positive hybrid quasiatom, and $\nu$ is the $E$-specification of the same-base negative hybrid quasiatom. Then $r_\nu^*$ and $r_\pi^*$ have the same canonical tuple, and $r_\nu^* \leq r_\pi^*$.*

To verify this claim, assume its conditions, and assume the hybrid quasiatoms that are $E$-specified by $\pi$ and $\nu$ are $P_q$-based. Let $(\vec{x})$ be the canonical tuple of $P^*$. We may assume that $(\vec{x}) = (\vec{y}, \vec{z})$,

---

[6]Of course, from the very beginning we could have chosen $r_\gamma$ to be $m$-ary rather than $n$-ary. But we were lazy to explain what $m$ was at the time when the arity of $r_\gamma$ was chosen. This could also have made clumsy some of our forthcoming definitions.

where $\vec{y}, \vec{z}$ are as in the paragraph following (22), and so is $\vec{c}$. Then, according to the stipulation of that paragraph, $r_\pi^* = r_\pi^*(\vec{y},\vec{z}) = \langle \Gamma^\pi \rangle \downarrow P^*(\vec{y},\vec{c})$ and $r_\nu^* = r_\nu^*(\vec{y},\vec{z}) = \langle \neg\Gamma^\nu \rangle \downarrow P^*(\vec{y},\vec{c})$. But, by condition 2 of Definition 3.14, $\Gamma^\pi$ is a $\top$-delay of $\neg\Gamma^\nu$. Therefore, as $P^*(\vec{y},\vec{z})$ and hence $P^*(\vec{y},\vec{c})$ is a static game, $\langle \neg\Gamma^\nu \rangle \downarrow P^*(\vec{y},\vec{c}) \leq \langle \Gamma^\pi \rangle \downarrow P^*(\vec{y},\vec{c})$. Thus, $r_\nu^* \leq r_\pi^*$, and Claim 1 is proven.

Let $E_1$ denote the result of replacing in $E$: 1) every $\sqcap$- and $\sqcap\!\!\!\sqcap$- (resp. $\sqcup$- and $\sqcup\!\!\!\sqcup$-) quasiatom by $\top$ (resp. $\bot$); 2) every surface occurrence $\gamma$ of each general atom by $r_\gamma$; and 3) every surface occurrence $\gamma$ of each hybrid letter[7] by $r_\gamma$; From (21), (22) and (23), applying Lemma 3.26 as many times as the number of non-elementary quasiatoms of $E$, we (with Lemma 3.12 in mind) get $\langle \Gamma \rangle \downarrow E^* = \langle \rangle \downarrow E_1^*$. But $E_1^*$ is an elementary game, and obviously for every elementary game $A$, $\langle \rangle \downarrow A = A$. Hence,

$$\langle \Gamma \rangle \downarrow E^* = E_1^*. \tag{24}$$

Assume $E$ has $m$ negative and $n$ positive general quasiatoms, $E$-specified by $\delta_1, \ldots, \delta_m$ and $\sigma_1, \ldots, \sigma_n$, respectively. Let $E_2$ be the result of replacing in $E_1$ each atom $r_{\delta_i}$ ($1 \leq i \leq m$) by $\top$, and each atom $r_{\sigma_i}$ ($1 \leq i \leq n$) by $\bot$. Applying Lemma 3.21 $m+n$ times, we get

$$E_2^* \leq E_1^*. \tag{25}$$

Next, assume $E$ contains $k$ hybrid letters, whose positive occurrences are $E$-specified by $\pi_1, \ldots, \pi_k$, and whose negative occurrences $E$-specified by $\nu_1, \ldots, \nu_k$, respectively. Let $E_3$ be the result of replacing in $E_2$ each letter $r_{\pi_i}$ ($1 \leq i \leq k$) by $r_{\nu_i}$. Applying Claim 1 plus Lemma 3.21 $k$ times, we get

$$E_3^* \leq E_2^*. \tag{26}$$

Let $q_1, \ldots, q_k$ be the list of the elementary components of the hybrid letters of $E$ whose negative occurrences are $E$-specified by the above $\nu_1, \ldots, \nu_k$, respectively. Compare $E_3$ with $\|E\|$. An analysis of how these two formulas have been obtained from $E$ can reveal that $E_3$ is just the result of replacing in $\|E\|$ every occurrence of every letter $q_i$ ($1 \leq i \leq k$) by $r_{\nu_i}$. That is, $E_3$ is a substitutional instance (in the standard classical sense) of $\|E\|$. The latter is classically valid because, by our assumptions, $E$ is stable. Hence $E_3$ is also classically valid, and thus $E_3^*$ is true. Then, by (26), (25) and (24), so is $\langle \Gamma \rangle \downarrow E^*$. This means nothing but that $\mathbf{Wn}^{E^*}\langle \Gamma \rangle = \top$. $\square$

## 4 Soundness of CL4

**Lemma 4.1** *If $\mathbf{CL4} \vdash F$, then $F$ is valid (any formula $F$). Moreover, there is an effective procedure that takes a $\mathbf{CL4}$-proof of an arbitrary formula $F$ and constructs an HPM that wins $F^*$ for every $F$-admissible interpretation $^*$.*

**Proof.** In view of Proposition 3.2 of [6], it would be sufficient to prove the above lemma — in particular, the 'Moreover' clause of it — with "fair EPM" instead of "HPM". Also, Lemmas 3.1 and 3.3 allow us to safely replace "$\mathbf{CL4}$-proof" by "reasonable $\mathbf{CL4}^\circ$-proof" in our present lemma. With these remarks in mind, Lemma 4.1 is an immediate consequence of the following Lemma 4.2. $\square$

**Lemma 4.2** *There is an effective procedure that takes a reasonable $\mathbf{CL4}^\circ$-proof of a hyperformula $F$ and constructs a fair EPM $\mathcal{E}$ such that, for every $F$-admissible interpretation $^*$, $\mathcal{E} \models F^*$.*

**Idea.** Every reasonable $\mathbf{CL4}^\circ$-proof, in fact, can be viewed as an interpretation-independent winning strategy for $\top$, and the fair EPM $\mathcal{E}$ that we are going to design just follows such a strategy. As we probably remember from the soundness proof given in [6], the same was the case with $\mathbf{CL3}$-proofs, where each conclusion-to-premise transition of Rule **A** encoded a move by $\bot$ (with all premises accounting for all possible legal moves by $\bot$), and the conclusion-to-premise transition of Rules **B1** and **B2** encoded the "good" move that ⊤ should make in a given situation; in either case, after a move was made, ⊤'s strategy would jump to the corresponding premise $H$, recursively calling itself on $H$. In our present case this intuitive meaning of Rules **A**, **B1** and **B2** is retained. In addition, Rule **C**° signals ⊤ that from now on it should try — using copy-cat methods — to keep identical[8] the subplays/subruns in the two occurrences of the hybrid atom introduced (in the bottom-up view) by that rule.

---

[7]Note that while in clause 2) the whole *atom* is replaced, in clause 3) only the *letter* is replaced, without affecting the attached tuple of terms of the atom.

The overall situation with **CL3**, however, was much simpler than it is with **CL4**°. In **CL3**-proof-derived strategies, as we just noted, to every legal move in a play corresponded a transition from a given formula to one of its premises $H$ in the proof. This is no longer the case with **CL4**°. Specifically, there is nothing in **CL4**°-proofs corresponding to moves made in non-choice quasiatoms. So, by the time when the strategy jumps to $H$, the game to which the original game will have been "brought down" may be not $H^*$ but rather $\langle\Omega\rangle H^*$, where $\Omega$ is the sequence of the moves made by the two players in the hybrid and general quasiatoms of $H$. Thus, the strategy has to be successful for such $\langle\Omega\rangle H^*$ rather than (as was the case with **CL3**-proof-derived strategies) just for $H^*$. Fortunately, it turns out that success in this more complicated situation is still possible as long as $\Omega$ is $H$-manageable; and ensuring that $\Omega$ is indeed always manageable also turns out to be a "manageable" task for ⊤. This is where all of our manageability-related lemmas from the previous section come to help.

Now a little more detailed — yet informal — description of how our strategy/machine $\mathcal{E}$ works for a hyperformula $F$ with a reasonable **CL4**°-proof. For safety and without loss of generality, we can limit our considerations to perfect interpretations and also assume that $F$ is closed. As noted, the strategy is recursive, at every step dealing with $\langle\Omega\rangle E^*$, where $E$ is a **CL4**°-provable reasonable hyperformula and $\Omega$ is an $E$-manageable (legal) position of $E^*$. Initially $E = F$ and $\Omega = \langle\rangle$. How $\mathcal{E}$ acts on $\langle\Omega\rangle E^*$ depends on by which of the four rules $E$ is derived in **CL4**°.

If $E$ is derived by Rule **B1** or **B2** from $H$, the machine — exactly as in [6] — makes the move $\alpha$ "prescribed" by that application of the rule. Say, if $E = G(7) \to \bigsqcup z G(z)$ and $H = G(7) \to G(7)$, then '2.7' is such a move. Lemmas 3.16 and 3.17 tell us that $\Omega$ remains $H$-manageable and that $\alpha$ brings $\langle\Omega\rangle E^*$ down to $\langle\Omega\rangle H^*$. So, after making move $\alpha$, the machine switches to its winning strategy for $\langle\Omega\rangle H^*$. This, by the induction hypothesis, guarantees a success.

If $E$ is derived by Rule **C**° from $H$ through replacing the two occurrences of a hybrid letter $P_q$ in $H$ by $P$, then the machine finds within $\Omega$ and copies, in the positive occurrence of $P_q$, all of the moves made so far by the environment in the negative occurrence of $P_q$ (or rather in the corresponding occurrence of the $P$-based atom), and vice versa. This series of moves brings the game down to $\langle\Omega'\rangle E^* = \langle\Omega'\rangle H^*$, where $\Omega'$ is the result of adding those moves to $\Omega$. Lemma 3.19 guarantees that $\Omega'$ is $H$-manageable. So, now the machine switches to its successful strategy for $\langle\Omega'\rangle H^*$ and eventually wins.

Finally, suppose $E$ is derived by Rule **A**. Our machine keeps granting permission. Now and then the environment will be making moves in general quasiatoms of $E$, to which $\mathcal{E}$ does not react. However, every time ⊥ makes a move in one of the hybrid quasiatoms, $\mathcal{E}$ copies that move in the other same-base quasiatom. Clauses (i) and (ii) of Lemma 3.20 guarantee that, while this is going on, (the continuously updated) $\Omega$ remains $E$-manageable. So, if nothing else happens, in view of Lemma 3.15, $\Omega$ — even if its grows infinite — remains $E$-manageable, and then Lemma 3.27 guarantees that the game will be won by ⊤ because, as a conclusion of Rule **A**, $E$ is stable. However, what will typically happen during this stage (except one — the last — case) is that sooner or later ⊥ makes a legal move *not* in a hybrid or general quasiatom, but rather a move signifying a decision associated with a choice quasiatom of $E$. E.g., if $E = (P_q \vee \neg P_q) \wedge (G_3 \sqcap G_4)$, then '2.1' can be such a move. Now the situation is very similar to the case with Rules **B1** or **B2**: the machine simply switches to its winning strategy for $\langle\Omega\rangle H^*$, where $H$ is the corresponding premise of $E$ ($H = (P_q \vee \neg P_q) \wedge G_3$ in our example). Clauses (iii)(2) and (iv)(2) of Lemma 3.20 guarantee that $\langle\Omega\rangle H^*$ is indeed the game to which $\langle\Omega\rangle E^*$ has evolved; and, according to clauses (iii)(1) and (iv)(1) of the same lemma, $\Omega$ is $H$-manageable, so that, by the induction hypothesis, $\mathcal{E}$ knows how to win $\langle\Omega\rangle H^*$.

**Proof.** Fix a hyperformula $F$ together with a reasonable **CL4**°-proof for it. In the present context we view such a proof as a sequence (rather than tree) of hyperformulas. We will be referring to this sequence as "the proof", and referring to hyperformulas occurring in the proof as "proof hyperformulas". We assume that there are no repetitions or other redundancies in the proof (otherwise eliminate them), and that each proof hyperformula comes with a fixed *justification* — an indication of by which rule and from what premises the hyperformula was derived.

---
[8]This is generally impossible in the literal sense, but what *is* possible is to ensure that one play is a ⊤-delay of the other which, taking into account that we are talking about static games, is just as good as if the two plays were fully identical.

We construct the EPM $\mathcal{E}$ whose work can be described as follows. At the beginning it creates three records: $E$ to hold a hyperformula, $\Omega$ to hold a position, and $f$ to hold a (representation of a) finite valuation.[9] $E$ is initialized to $F$, $\Omega$ initialized to $\langle\rangle$, and $f$ initialized to $\{x_1/c_1, \ldots, x_s/c_s\}$, where $x_1, \ldots, x_s$ are all the free variables of $F$ and, for each $1 \le i \le s$, $c_i$ is the value assigned to $x_i$ by the valuation spelled on the valuation tape. After this initialization step, $\mathcal{E}$ follows the following interactive algorithm MAIN LOOP. The description of this algorithm assumes that, at the beginning of each iteration of MAIN LOOP or INNER LOOP, $E$ is a proof hyperformula, and also that, for every free variable $x$ of $E$, $f$ contains exactly one element $x/c$ and vice versa: for every element $x/c$ of $f$, $x$ is a free variable of $E$. That these conditions are always satisfied can be immediately seen from the description of the algorithm.

**Procedure** MAIN LOOP: Act depending on which of the four rules was used (last) to derive $E$ in the proof:

**Case of Rule B1** Let $H$ be the premise of $E$ in the proof. $H$ is the result of replacing in $E$ a certain negative (resp. positive) surface occurrence $\gamma$ of a subhyperformula $G_1 \sqcap \ldots \sqcap G_n$ (resp. $G_1 \sqcup \ldots \sqcup G_n$) by $G_i$ for some $i \in \{1, \ldots, n\}$. Then make the move $\gamma i$; update $f$ by deleting in it all pairs $z/c$ where $z$ is not a free variable of $H$; update $E$ to $H$; repeat MAIN LOOP.

**Case of Rule B2** Let $H$ be the premise of $E$ in the proof. $H$ is the result of replacing in $E$ a negative (resp. positive) surface occurrence $\gamma$ of a subhyperformula $\sqcap xG(x)$ (resp. $\sqcup xG(x)$) by $G(t)$ for some term $t$ such that (if $t$ is a variable) neither the above occurrence of $\sqcap xG(x)$ (resp. $\sqcup xG(x)$) in $E$ nor any of the free occurrences of $x$ in $G$ are in the scope of $\forall t, \exists t, \sqcap t$ or $\sqcup t$. Let $c = f(t)$.[10] Then make the move $\gamma c$; as long as $t$ is a free variable of $H$, update $f$ to $f \cup \{t/c\}$; update $E$ to $H$; repeat MAIN LOOP.

**Case of Rule C°** Let $H$ be the premise of $E$ in the proof. $E$ is the result of replacing in $H$ the positive surface occurrence $\pi$ and the negative surface occurrence $\nu$ of some hybrid letter $P_q$ by the general letter $P$. Let $\langle \bot\pi_1, \ldots, \bot\pi_n \rangle$ and $\langle \bot\nu_1, \ldots, \bot\nu_m \rangle$ be $\Omega^\pi$ and $\Omega^\nu$, respectively. Then: make the $n + m$ moves $\pi\nu_1, \ldots, \pi\nu_m, \nu\pi_1, \ldots, \nu\pi_n$ (in this very order); update $\Omega$ to $\langle \Omega, \top\pi\nu_1, \ldots, \top\pi\nu_m, \top\nu\pi_1, \ldots, \top\nu\pi_n \rangle$; update $E$ to $H$; repeat MAIN LOOP.

**Case of Rule A** Follow the procedure INNER LOOP described below.

> **Procedure** INNER LOOP: Keep granting permission until the adversary makes a move $\alpha$, then act depending on which of the following five subcases holds:
>
> **Subcase (i)** $\alpha = \gamma\beta$, where $\gamma$ $E$-specifies a general quasiatom. Then update $\Omega$ to $\langle \Omega, \bot\gamma\beta \rangle$, and repeat INNER LOOP.
>
> **Subcase (ii)** $\alpha = \gamma\beta$, where $\gamma$ $E$-specifies a hybrid quasiatom. Let $\sigma$ be the $E$-specification of the other same-base hybrid quasiatom. Then make the move $\sigma\beta$, update $\Omega$ to $\langle \Omega, \bot\gamma\beta, \top\sigma\beta \rangle$, and repeat INNER LOOP.
>
> **Subcase (iii)** $\alpha = \gamma i$, where $\gamma$ $E$-specifies a positive (resp. negative) quasiatom $G_1 \sqcap \ldots \sqcap G_n$ (resp. $G_1 \sqcup \ldots \sqcup G_n$) and $i \in \{1, \ldots, n\}$. Let $H$ be the result of replacing in $E$ the above quasiatom by $G_i$. Then update $f$ by deleting in it all pairs $z/c$ where $z$ is not a free variable of $H$; update $E$ to $H$; repeat MAIN LOOP.
>
> **Subcase (iv)** $\alpha = \gamma c$, where $\gamma$ $E$-specifies a positive (resp. negative) quasiatom $\sqcap xG(x)$ (resp. $\sqcup xG(x)$). Let $H$ be premise of $E$ in the proof that is the result of replacing in $E$ the above quasiatom by $G(y)$ for some variable $y$ not occurring in $E$. Then, as long as $y$ is a free variable of $H$,[11] update $f$ to $f \cup \{y/c\}$; update $E$ to $H$; repeat MAIN LOOP.

---

[9]Remember from [6] that a finite valuation $f$ is one that sends only a finite number of variables to something other than 0. We agreed to represent such an $f$ as a finite set of variable/constant pairs that has to include an entry for each variable $z$ with $f(z) \ne 0$, and may or may not include entries for other variables.

[10]Remember that, if $t$ is a constant, then $f$ (as any other valuation) sends it to itself, and if $t$ is a variable not listed within the variable/constant pairs of the representation of $f$, then $f(t) = 0$.

[11]I.e. unless $x$ did not really have a free occurrence in $G(x)$.

**Subcase (v)** $\alpha$ does not satisfy the conditions of any of the above Subcases (i)-(iv). Then go to an infinite loop in a permission state.

It is obvious that $\mathcal{E}$ can be constructed effectively from the **CL4**°-proof of $F$. Pick an arbitrary $F$-admissible interpretation $^*$, an arbitrary valuation $e$ and an arbitrary $e$-computation branch $B$ of $\mathcal{E}$. Fix $\Theta_\infty$ as the run spelled by $B$. Our goal is to show that $B$ is fair and that $\mathbf{Wn}_e^{F^*}\langle\Theta_\infty\rangle = \top$.

Consider the work of $\mathcal{E}$ in $B$. For each $k \geq 1$ such that MAIN LOOP makes at least $k$ iterations in $B$, let $E_k$ and $f_k$ denote the values of records $E$ and $f$ at the beginning of the $k$th iteration of MAIN LOOP, and let $K_k = f_k E_k$. Evidently $E_{k+1}$ is always one of the premises of $E_k$ in the proof, so that MAIN LOOP is iterated only a finite number of times. Fix $l$ as the number of these iterations. The $l$th (last) iteration deals with the case of Rule **A** — and, besides, never with Subcases (iii) or (iv) within it — for otherwise there would be a next iteration of MAIN LOOP. This guarantees that $\mathcal{E}$ will grant permission infinitely many times during the $l$th iteration, so that branch $B$ is indeed fair.

What remains to show now is that $\mathbf{Wn}_e^{F^*}\langle\Theta_\infty\rangle = \top$, i.e. $\mathbf{Wn}^{e[F^*]}\langle\Theta_\infty\rangle = \top$. $f_1$ agrees with $e$ on all free variables of $F = E_1$. Hence, by Lemma 3.7, $e[F^*] = e[(f_1 E_1)^*] = e[K_1^*]$. Let $^\star$ be the perfect interpretation induced by $(^*, e)$. By Lemma 3.9, $e[K_1^*] = K_1^\star$. Hence $e[F^*] = K_1^\star$. Thus, our goal of showing $\mathbf{Wn}^{e[F^*]}\langle\Theta_\infty\rangle = \top$ now reduces to showing $\mathbf{Wn}^{K_1^\star}\langle\Theta_\infty\rangle = \top$. The latter is immediate when $\Theta_\infty$ is a $\bot$-illegal run of $K_1^\star$. Hence we exclude this trivial case and, for the rest of this proof, assume that $\Theta_\infty$ is not a $\bot$-illegal run of $K_1^*$. Speaking less formally, we assume that $\bot$ never makes illegal moves. Note that the $K_i^\star$ are constant games (Lemma 3.6), so we can talk about their legal, won, etc. runs — as we just did — without mentioning $e$ or bothering about the distinction between "legal" and "unilegal".

The above-observed fact that $E_l$ is derived by Rule **A** implies that

$$E_l \text{ is stable.} \tag{27}$$

For each $i$ with $1 \leq k \leq l$, let $\Theta_k$ be the initial segment of $\Theta_\infty$ consisting of the moves made by the beginning of the $k$th iteration of MAIN LOOP. Aslo, for each such $k$, let $\Omega_k$ be the value of record $\Omega$ at the beginning of the $k$th iteration of MAIN LOOP.

**Claim 1** *For any $k$ with $1 \leq k \leq l$,*

$$\Omega_k \text{ is } K_k\text{-manageable;} \tag{28}$$
$$\langle\Theta_k\rangle K_1^\star = \langle\Omega_k\rangle K_k^\star. \tag{29}$$

This claim can be proven by induction on $k$. The basis case with $k = 1$ is trivial as $\Theta_1 = \Omega_1 = \langle\rangle$.

Now consider an arbitrary $k$ with $1 \leq k < l$ and assume (induction hypothesis) that conditions (28)-(29) are satisfied. Note that, by Convention 3.10, condition (29) implies $\Theta_k \in \mathbf{LR}^{K_1^\star}$ and $\Omega_k \in \mathbf{LR}^{K_k^\star}$. Implicitly relying on this fact, we separately consider the following four cases, depending on with which case the $k$th iteration of MAIN LOOP deals. In each case we want to show that the same two conditions continue to be satisfied for $k + 1$, i.e. that the following statements are true:

$$\Omega_{k+1} \text{ is } K_{k+1}\text{-manageable;} \tag{30}$$
$$\langle\Theta_{k+1}\rangle K_1^\star = \langle\Omega_{k+1}\rangle K_{k+1}^\star. \tag{31}$$

*Case of Rule* **B1**. Record $\Omega$ is not updated in this case, so $\Omega_{k+1} = \Omega_k$. Also, exactly one ($\top$-labeled) move $\gamma i$ is made, where $\gamma$ is the $E_k$- and hence $K_k$-specification of a negative (resp. positive) quasiatom $G_1 \sqcap \ldots \sqcap G_n$ (resp. $G_1 \sqcap \ldots \sqcap G_n$) and $i \in \{1, \ldots, n\}$. Thus, $\Theta_{k+1} = \langle\Theta_k, \top\gamma i\rangle$. Note that $K_{k+1}$ relates to $K_k$ as $H$ does to $E$ in Lemma 3.16. Keeping this in mind, with $\Omega_k = \Omega_{k+1}$ in the role of $\Omega$ (which, by our induction hypothesis, is a $K_k$-manageable unilegal run of $K_k$), we find that (30) follows from clause 1 of Lemma 3.16. According to clause 2 of the same lemma, $\langle\Omega_k, \top\gamma i\rangle K_k^\star = \langle\Omega_k\rangle K_{k+1}^\star = \langle\Omega_{k+1}\rangle K_{k+1}^\star$, i.e. $\langle\top\gamma i\rangle\langle\Omega_k\rangle K_k^\star = \langle\Omega_{k+1}\rangle K_{k+1}^\star$.[12] But, by (29), $\langle\Omega_k\rangle K_k^\star = \langle\Theta_k\rangle K_1^\star$. Hence, $\langle\top\gamma i\rangle\langle\Theta_k\rangle K_1^\star = \langle\Omega_{k+1}\rangle K_{k+1}^\star$, i.e. $\langle\Theta_k, \top\gamma i\rangle K_1^\star = \langle\Omega_{k+1}\rangle K_{k+1}^\star$, which, as $\langle\Theta_k, \top\gamma i\rangle = \Theta_{k+1}$, proves (31).

---

[12] Here and very often later we implicitly rely on the fact that, for any game $A$ and positions $\Psi_1$ and $\Psi_2$, the conditions "$\langle\Psi_1, \Psi_2\rangle \in \mathbf{LR}^A$" and "$\Psi_1 \in \mathbf{LR}^A$ and $\Psi_2 \in \mathbf{LR}^{\langle\Psi_1\rangle A}$" are equivalent and, whenever they are satisfied, we have $\langle\Psi_1, \Psi_2\rangle A = \langle\Psi_2\rangle\langle\Psi_1\rangle A$. ($\langle\Psi_2\rangle\langle\Psi_1\rangle A$ should be read as $\langle\Psi_2\rangle(\langle\Psi_1\rangle A)$.) An official proof of this easy-to-understand phenomenon is given in [5], Lemma 3.4.

*Case of Rule* **B2**. Similar to the previous one, only use Lemma 3.17 instead of Lemma 3.16.

*Case of Rule* **C°**. With $E_k$ in the role of $E$, let $H$, $\pi$, $\nu$, $\pi_1,\ldots,\pi_n$, $\nu_1,\ldots,\nu_m$ be as in the description of the 'Case of Rule **C°**' clause of MAIN LOOP. Note that $E_{k+1} = H$ and

$$\Omega_{k+1} = \langle \Omega_k, \top\pi\nu_1, \ldots, \top\pi\nu_m, \top\nu\pi_1, \ldots, \top\nu\pi_n \rangle.$$

Now, with $\Omega_k$ in the role of $\Omega$, $K_k$ in the role of $E$ and $K_{k+1}$ in the role of $H$, the conditions of Lemma 3.19 are satisfied. Then, by that lemma, $\Omega_{k+1}$ is a $K_{k+1}$-manageable (uni)legal position of $K_{k+1}^\star$, which proves (30) and also ensures that $\langle \Omega_{k+1}\rangle K_{k+1}^\star$ is defined. The latter can be rewritten as

$$\langle \top\pi\nu_1, \ldots, \top\pi\nu_m, \top\nu\pi_1, \ldots, \top\nu\pi_n \rangle \langle \Omega_k \rangle K_{k+1}^\star.$$

By (29), $\langle \Omega_k \rangle K_k^\star = \langle \Theta_k \rangle K_1^\star$. Also notice that $K_{k+1}^\star = K_k^\star$, so $\langle \Omega_k \rangle K_{k+1}^\star = \langle \Theta_k \rangle K_1^\star$. Hence $\langle \Omega_{k+1}\rangle K_{k+1}^\star = \langle \top\pi\nu_1, \ldots, \top\pi\nu_m, \top\nu\pi_1, \ldots, \top\nu\pi_n \rangle \langle \Theta_k \rangle K_1^\star$. This means that

$$\langle \Omega_{k+1}\rangle K_{k+1}^\star = \langle \Theta_k, \top\pi\nu_1, \ldots, \top\pi\nu_m, \top\nu\pi_1, \ldots, \top\nu\pi_n \rangle K_1^\star.$$

But $\Theta_{k+1} = \langle \Theta_k, \top\pi\nu_1, \ldots, \top\pi\nu_m, \top\nu\pi_1, \ldots, \top\nu\pi_n \rangle$. Hence $\langle \Omega_{k+1}\rangle K_{k+1}^\star = \langle \Theta_{k+1}\rangle K_1^\star$. This proves (31).

*Case of Rule* **A**. The work of the machine during this ($k$th) iteration of MAIN LOOP consists of iterating INNER LOOP a finite number $l_k$ of times (otherwise we would have $k = l$). Fix this $l_k$. For each $m$ with $1 \leq m \leq l_k$, let $\Theta_{k,m}$ be the initial segment of $\Theta_\infty$ consisting of the moves made by the beginning of the $m$th iteration of INNER LOOP within the $k$th iteration of MAIN LOOP. Similarly, for each such $m$, let $\Omega_{k,m}$ be the value of record $\Omega$ at the beginning of the $m$th iteration of INNER LOOP within the $k$th iteration of MAIN LOOP.

**Subclaim 1.1** *For any $m$ with $1 \leq m \leq l_k$,*

$$\Omega_{k,m} \text{ is } K_k\text{-manageable;} \tag{32}$$

$$\langle \Theta_{k,m}\rangle K_1^\star = \langle \Omega_{k,m}\rangle K_k^\star. \tag{33}$$

We prove the above by induction on $m$. The basis with $m = 1$ is straightforward: we have $\Theta_{k,1} = \Theta_k$ and $\Omega_{k,1} = \Omega_k$, so this case is nothing but the induction hypothesis of the induction on $k$ in our proof of Claim 1, i.e. statements (28)-(29).

Now consider any $m$ with $1 \leq m < l_k$, and assume (32) and (33) are true (induction hypothesis). Let us take a note that, by (33) and Convention 3.10, we have $\Theta_{k,m} \in \mathbf{LR}^{K_1^\star}$ and $\Omega_{k,m} \in \mathbf{LR}^{K_k^\star}$. We want to show that the following conditions are also satisfied:

$$\Omega_{k,m+1} \text{ is } K_k\text{-manageable;} \tag{34}$$

$$\langle \Theta_{k,m+1}\rangle K_1^\star = \langle \Omega_{k,m+1}\rangle K_k^\star. \tag{35}$$

At the beginning of the $m$th iteration of INNER LOOP, the machine is waiting for the adversary to make a move $\alpha$. Such a move must be made because $k$ is not the last iteration of MAIN LOOP. By our assumption that $\bot$ never makes illegal moves, $\langle \Theta_{k,m}, \bot\alpha \rangle$ must be a legal position of $K_1^\star$, whence $\langle \bot\alpha\rangle \in \mathbf{LR}^{\langle \Theta_{k,m}\rangle K_1^\star}$, whence, by (33), $\langle \bot\alpha\rangle \in \mathbf{LR}^{\langle \Omega_{k,m}\rangle K_k^\star}$, whence $\langle \Omega_{k,m}, \bot\alpha\rangle \in \mathbf{LR}^{K_k^\star}$. This means that $\alpha$, $K_k$ (in the role of $E$) and $\Omega_{k,m}$ (in the role of $\Omega$) satisfy the conditions of Lemma 3.20. Then, in view of that lemma, it is obvious that $\alpha$ — now with $E_k$ in the role of $E$ — should satisfy the conditions of one of the Subcases **(i)**, **(ii)**, **(iii)** or **(iv)** from the description of INNER LOOP. Subcases **(iii)** and **(iv)** are impossible because then we would have $m = l_k$. Thus, either Subcase **(i)** or Subcase **(ii)** should be satisfied.

Suppose Subcase **(i)** is satisfied, with $\alpha = \gamma\beta$ as described in that subcase. Then $\Omega_{k,m+1} = \langle \Omega_{k,m}, \bot\gamma\beta\rangle$, and (34) holds by clause (i) of Lemma 3.20, which also asserts that $\langle \Omega_{k,m}, \bot\gamma\beta\rangle \in \mathbf{LR}^{K_k^\star}$. According to (33), $\langle \Omega_{k,m}\rangle K_k^\star = \langle \Theta_{k,m}\rangle K_1^\star$, and hence $\langle \Omega_{k,m}, \bot\gamma\beta\rangle K_k^\star = \langle \Theta_{k,m}, \bot\gamma\beta\rangle K_1^\star$, i.e. $\langle \Omega_{k,m+1}\rangle K_k^\star = \langle \Theta_{k,m}, \bot\gamma\beta\rangle K_1^\star$. But in our case $\Theta_{k,m+1} = \langle \Theta_{k,m}, \bot\gamma\beta\rangle$. Therefore, $\langle \Theta_{k,m+1}\rangle K_1^\star = \langle \Omega_{k,m+1}\rangle K_k^\star$. This proves (35).

Suppose now Subcase **(ii)** is satisfied, with $\alpha = \gamma\beta$ and $\sigma$ as described in that subcase. Then $\Omega_{k,m+1} = \langle \Omega_{k,m}, \bot\gamma\beta, \top\sigma\beta\rangle$. Now, (34) follows from clause (ii) of Lemma 3.20, which also asserts that $\langle \Omega_{k,m}, \bot\gamma\beta, \top\sigma\beta\rangle$ is in $\mathbf{LR}^{K_k^\star}$. And, taking into account that $\Theta_{k,m+1} = \langle \Theta_{k,m}, \bot\gamma\beta, \top\sigma\beta\rangle$, (35) follows from (33). Subclaim 1.1 is proven.

Back to the 'Case of Rule **A**' step of our proof of Claim 1. Consider the $l_k$th (last) iteration of INNER LOOP within the $k$th iteration of MAIN LOOP. Since $k < l$, obviously this iteration deals with one of the Subcases **(iii)** or **(iv)**. Let $\alpha$ and $H$ be as in the description of Subcase **(iii)** (resp. **(iv)**). Note that $E_{k+1} = H$ and $\Omega_{k+1} = \Omega_{k,l_k}$. For the same reasons as in the proof of Subclaim 1.1 (with $l_k$ in the role of $m$), $\langle \Omega_{k,l_k}, \bot\alpha\rangle$ is a legal position of $K_k^\star$. According to Subclaim 1.1, we have:

$$\Omega_{k,l_k} \text{ is } K_k\text{-manageable;} \tag{36}$$
$$\langle \Theta_{k,l_k}\rangle K_1^\star = \langle \Omega_{k,l_k}\rangle K_k^\star. \tag{37}$$

Statement (30) follows from (36) by clause (iii)(1) (resp. (iv)(1)) of Lemma 3.20. According to clause (iii)(2) (resp. (iv)(2)) of the same lemma, $\langle\Omega_{k,l_k},\bot\alpha\rangle K_k^\star = \langle\Omega_{k,l_k}\rangle K_{k+1}^\star$ and hence $\langle\Omega_{k,l_k},\bot\alpha\rangle K_k^\star = \langle\Omega_{k+1}\rangle K_{k+1}^\star$. This, in turn, implies $\langle\bot\alpha\rangle\langle\Omega_{k,l_k}\rangle K_k^\star = \langle\Omega_{k+1}\rangle K_{k+1}^\star$. By (37), $\langle\Omega_{k,l_k}\rangle K_k^\star = \langle\Theta_{k,l_k}\rangle K_1^\star$. Hence $\langle\bot\alpha\rangle\langle\Theta_{k,l_k}\rangle K_1^\star = \langle\Omega_{k+1}\rangle K_{k+1}^\star$ and thus $\langle\Theta_{k,l_k},\bot\alpha\rangle K_1^\star = \langle\Omega_{k+1}\rangle K_{k+1}^\star$. But observe that $\langle\Theta_{k,l_k},\bot\alpha\rangle = \Theta_{k+1}$. Therefore (31) holds. Claim 1 is proven.

We continue our proof of Lemma 4.2. Consider the last ($l$th) iteration of MAIN LOOP. As we noted earlier when deriving (27), this iteration deals with the case of Rule **A**. Let $\mathcal{N}$ be $\{1,\ldots,k\}$ if $k$ is the number of iterations of INNER LOOP within the $l$th iteration of MAIN LOOP, and be $\{1,2,3,\ldots\}$ if there are infinitely many such iterations. For each $m \in \mathcal{N}$, as before, let $\Theta_{l,m}$ be the initial segment of $\Theta_\infty$ consisting of the moves made by the beginning of the $m$th iteration of INNER LOOP within the $l$th iteration of MAIN LOOP, and let $\Omega_{l,m}$ be the value of record $\Omega$ at the beginning of the $m$th iteration of INNER LOOP within the $l$th iteration of MAIN LOOP. For the same reasons[13] as in the proof of (32) and (33) (where from (33) we only need its implicit statement that $\Omega_{k,m} \in \mathbf{LR}^{K_k^\star}$), we have:

$$\text{For any } m \in \mathcal{N}, \ \Omega_{l,m} \text{ is a } K_l\text{-manageable legal position of } K_l^\star. \tag{38}$$

Note that, during the work of $\mathcal{E}$, every update of record $\Omega$ extends its previous value by adding new labmoves to it, without ever deleting old labmoves. So, let $\Omega_\infty$ be the "ultimate" value of $\Omega$, precisely meaning the shortest run such that, for every $m \in \mathcal{N}$, $\Omega_{l,m}$ is an initial segment of $\Omega_\infty$. Of course, if $\mathcal{N} = \{1,\ldots,k\}$, then $\Omega_\infty$ is simply $\Omega_{l,k}$. Statement (38) — together with Lemma 3.15 when $\Omega_\infty$ is infinite — implies that $\Omega_\infty$ is a $K_l$-manageable legal run of $K_l^\star$. Therefore, by (27) and Lemma 3.27, we have

$$\mathbf{Wn}^{K_l^\star}\langle\Omega_\infty\rangle = \top. \tag{39}$$

$\Omega_\infty$ is an extension of $\Omega_l$, so that $\Omega_\infty = \langle\Omega_l, \Delta\rangle$ for some run $\Delta$. Let us fix this $\Delta$. Now (39) can be rewritten as $\mathbf{Wn}^{K_l^\star}\langle\Omega_l,\Delta\rangle = \top$. In turn, the latter — remembering the definition of the operation of prefixation and taking into account that, by (29), $\Omega_l \in \mathbf{LR}^{K_l^\star}$ — can be rewritten as $\mathbf{Wn}^{\langle\Omega_l\rangle K_l^\star}\langle\Delta\rangle = \top$. Now, according to Claim 1, $\langle\Omega_l\rangle K_l^\star = \langle\Theta_l\rangle K_1^\star$. Thus, $\mathbf{Wn}^{\langle\Theta_l\rangle K_1^\star}\langle\Delta\rangle = \top$, which can be rewritten back as

$$\mathbf{Wn}^{K_1^\star}\langle\Theta_l,\Delta\rangle = \top. \tag{40}$$

For the same reasons[14] as in our proof of Subclaim 1.1, iterations of INNER LOOP within the $l$th iteration of MAIN LOOP never deal with Subcases (iii), (iv) or (v) of 'Case of Rule **A**'. The remaining Subcases (i) and (ii) add to record $\Omega$ all of the moves made by the players (and no other moves, of course). Therefore, taking into account that the value of that record is $\Omega_l$ when the $l$th iteration of MAIN LOOP starts, we can see that $\Delta$ is nothing but exactly the sequence of all moves made during the $l$th iteration of MAIN LOOP. Hence $\Theta_\infty = \langle\Theta_l, \Delta\rangle$. Thus, by (40), $\mathbf{Wn}^{K_1^\star}\langle\Theta_\infty\rangle = \top$, and our proof of Lemma 4.2 is complete. □

---

[13] With the minor difference that now the reason why Subcases (iii) and (iv) are impossible is that otherwise $l$ would not be the last iteration of MAIN LOOP.

[14] Again, with the minor difference pointed out in the previous footnote.

# 5 Completeness of CL4

**Lemma 5.1** *If* **CL4** $\not\vdash F$, *then* $F$ *is not valid (any formula* $F$). *Moreover, if* **CL4** $\not\vdash F$, *then* $F^*$ *is not computable for some* $F$-*admissible interpretation* $^*$ *that interprets elementary letters as finitary predicates of arithmetical complexity* $\Delta_2$, *and interprets general letters as problems of the form* $(A_1^1 \sqcup \ldots \sqcup A_m^1) \sqcap \ldots \sqcap (A_m^1 \sqcup \ldots \sqcup A_m^m)$, *where each* $A_i^j$ *is a finitary predicate of arithmetical complexity* $\Delta_2$.

**Idea.** The reader may want to start by trying the following *Exercise*: Verify that $P \to P \land P$ is not valid. *Hint*: Show that **CL3** $\not\vdash (p_1 \sqcap p_2) \to (p_1 \sqcap p_2) \land (p_1 \sqcap p_2)$, then remember the completeness theorem for **CL3**.

The above hint provides insights into the basic idea underlying our proof of Lemma 5.1. We are going to show that, whenever **CL4** $\not\vdash F$, there is a **CL3**-formula $\lceil F \rceil$ of the same form as $F$ that is not provable in **CL3**; this, in view of the already known completeness of **CL3**, immediately yields non-validity for $F$. Precisely, "the same form as $F$" will mean that $\lceil F \rceil$ is the result of rewriting in $F$ every $n$-ary general atom $P(\vec{t})$ as $\bigl(\check{P}_1^1(\vec{t}) \sqcup \ldots \sqcup \check{P}_m^1(\vec{t})\bigr) \sqcap \ldots \sqcap \bigl(\check{P}_1^m(\vec{t}) \sqcup \ldots \sqcup \check{P}_m^m(\vec{t})\bigr)$, where $m$ is a "sufficiently big" number and the $\check{P}_j^i$ are arbitrary "neutral" (not occurring in $F$ and pairwise distinct) $n$-ary elementary letters. Let us call the above long formula a *molecule* and say that the general letter $P$ is its *base*.

Intuitively, the reason why **CL3** $\not\vdash \lceil F \rceil$, i.e. why $\top$ cannot win (the game represented by) $\lceil F \rceil$, is that a smart environment may start choosing different conjuncts/disjuncts in different occurrences of molecules. The best that $\top$ can do in such a play is to match any given positive or negative occurrence of a molecule with one (but not more!) negative or positive occurrence of a same-base molecule — match in the sense that mimic environment's moves in order to keep the subgames/subformulas at the two occurrences identical. Yet, this is not sufficient for $\top$ to achieve a guaranteed success. This is so because $\top$'s decisions about what pairs of molecules to match in $\lceil F \rceil$ can be modeled by appropriate applications of Rule **C** in an attempted **CL4**-proof of $F$, and so can be — through Rules **A**, **B1** and **B2** — either player's decisions required by the choice operators in the non-molecule parts of $\lceil F \rceil$. A winning strategy (**CL3**-proof) for $\lceil F \rceil$ would then translate into a **CL4**-proof for $F$, which, however, does not exist.

**Proof.** Fix a **CL4**-formula $F$. Let $\mathcal{P}$ be the set of all general letters occurring in $F$. Let us fix $m$ as the total number of occurrences of such letters in $F$;[15] if there are fewer than 2 of such occurrences, then we take $m = 2$. For the rest of this section, let us agree that

$$a, b \text{ always range over } \{1, \ldots, m\}.$$

Let us pick some pool $z_1, z_2, \ldots$ of variables not occurring in $F$. For each $n$-ary $P \in \mathcal{P}$ and each $a, b$, with $\vec{z}$ abbreviating $z_1, \ldots, z_n$, we fix the atom

- $\check{P}_b^a(\vec{z})$,

where $\check{P}_b^a$ is an $n$-ary non-logical elementary letter not occurring in $F$. We assume that $\check{P}_b^a \neq \check{Q}_d^c$ as long as either $P \neq Q$ or $a \neq c$ or $b \neq d$. Note that the $\check{P}_b^a$ are elementary letters despite our "tradition" according to which the capital letters $P, Q, \ldots$ stand for general letters.

Next, for each $n$-ary $P \in \mathcal{P}$ and each $a$, we define

- $\check{P}_{\sqcup}^a(\vec{z}) = \check{P}_1^a(\vec{z}) \sqcup \ldots \sqcup \check{P}_m^a(\vec{z})$.

Finally, for each $n$-ary $P \in \mathcal{P}$, we define

- $\check{P}_{\sqcup}^{\sqcap}(\vec{z}) = \check{P}_{\sqcup}^1(\vec{z}) \sqcap \ldots \sqcap \check{P}_{\sqcup}^m(\vec{z}) = \bigl(\check{P}_1^1(\vec{z}) \sqcup \ldots \sqcup \check{P}_m^1(\vec{z})\bigr) \sqcap \ldots \sqcap \bigl(\check{P}_1^m(\vec{z}) \sqcup \ldots \sqcup \check{P}_m^m(\vec{z})\bigr)$.

For an $n$-ary $P \in \mathcal{P}$ and an $n$-tuple $\vec{t}$ of terms, we refer to the formulas $\check{P}_b^a(\vec{t})$, $\check{P}_{\sqcup}^a(\vec{t})$ and $\check{P}_{\sqcup}^{\sqcap}(\vec{t})$ as **molecules**, in particular, $P(\vec{t})$-**based molecules**, where "$(\vec{t})$" can be omitted when there is no need to indicate this parameter. We can also more specifically refer to $\check{P}_b^a(\vec{t})$ as a[16] $\check{P}_b^a$-**molecule**, refer to $\check{P}_{\sqcup}^a(\vec{t})$ as a $\check{P}_{\sqcup}^a$-**molecule**, and refer to $\check{P}_{\sqcup}^{\sqcap}(\vec{t})$ as a $\check{P}_{\sqcup}^{\sqcap}$-**molecule**. To differentiate between the three sorts of molecules, we call the molecules of the type $\check{P}_b^a(\vec{t})$ **small**, call the molecules of the type $\check{P}_\sqcup^a(\vec{t})$ **medium**, and call the molecules of the type $\check{P}_\sqcup^\sqcap(\vec{t})$ **large**.

---

[15]In fact, a much smaller $m$ would be sufficient for our purposes. E.g., $m$ can be chosen to be such that no given general letter has more than $m$ occurrences in $F$. But why try to economize.

[16]We say "a" rather than "the" here because, while $P$, $a$ and $b$ are fixed, $\vec{t}$ may vary.

For simplicity, for the rest of this section we assume/pretend that the languages of **CL3** and **CL4** have no non-logical letters other than those occurring in $F$ plus the letters $\check{P}_b^a$ ($P \in \mathcal{P}$, $a, b \in \{1, \ldots, m\}$). This way the scopes of the terms "**CL4**-formula" and "**CL3**-formula" are correspondingly redefined.

Let us say that an occurrence of a molecule in a given **CL3**-formula is **independent** iff it is not a part of another ("larger") molecule. E.g., the occurrence of $\check{P}_b^a(\vec{t})$ in $\check{P}_b^a(\vec{t}) \to \bot$ is independent, while in $\check{P}_\sqcup^a(\vec{t}) \to \bot$, i.e. in $\check{P}_1^a(\vec{t}) \sqcup \ldots \sqcup \check{P}_b^a(\vec{t}) \sqcup \ldots \sqcup \check{P}_m^a(\vec{t}) \to \bot$, it is not. Of course, surface occurrences of molecules are always independent, and so are any — surface or non-surface — occurrences of large molecules.

We say that a **CL3**-formula $E$ is **good** iff the following conditions are satisfied:

**Cond1:** $E$ contains at most $m$ independent occurrences of molecules.

**Cond2:** Only large molecules (may) have independent non-surface occurrences in $E$.

**Cond3:** For each $a$ and $b$, $E$ has at most one positive independent occurrence of a $\check{P}_b^a$-molecule and at most one negative independent occurrence of a $\check{P}_b^a$-molecule.

**Cond4:** For each $a$, $E$ has at most one positive independent occurrence of a $\check{P}_\sqcup^a$-molecule, and when $E$ has such an occurrence, then for no $b$ does $E$ have a positive independent occurrence of a $\check{P}_b^a$-molecule.

Let $E$ be a **CL3**-formula. By an **isolated** small molecule of $E$ (or $E$**-isolated** small molecule, or a small molecule **isolated in** $E$) we will mean a $\check{P}_b^a$-molecule that has an independent occurrence in $E$, such that $E$ contains no other independent occurrences of $\check{P}_b^a$-molecules. We will say that such a molecule is **positive** or **negative** depending on whether its independent occurrence in $E$ is positive or negative. Next, the **floorification** of $E$, denoted by

$$\lfloor E \rfloor,$$

is the result of replacing in $E$ every independent occurrence of every $P(\vec{t})$-based large, medium and $E$-isolated small molecule by the general atom $P(\vec{t})$.

**Claim 1** *For any good* **CL3**-*formula* $E$, *if* **CL3** $\vdash E$, *then* **CL4** $\vdash \lfloor E \rfloor$.

To prove this claim, assume $E$ is good and **CL3** $\vdash E$. By induction on the length of the **CL3**-proof of $E$, we want to show that **CL4** $\vdash \lfloor E \rfloor$. We need to consider the following three cases, depending on which of the three rules of **CL3** was used (last) to derive $E$.

*CASE 1:* $E$ is derived by Rule **A**. Let us fix the set $\vec{H}$ of premises of $E$. Each element of $\vec{H}$ is provable in **CL3**. Hence, by the induction hypothesis, we have:

$$\text{For any } H \in \vec{H}, \text{ if } H \text{ is good, then } \mathbf{CL4} \vdash \lfloor H \rfloor. \tag{41}$$

We consider the following 3 subcases. The first two subcases are not mutually exclusive, and either can be chosen when both apply. Specifically, Subcase 1.1 (resp. 1.2) is about when $E$ has a positive (resp. negative) surface occurrence of a large (resp. medium) molecule. Then, as we are going to see, replacing that molecule by a "safe" conjunct (resp. disjunct), corresponding to a smart environment's possible move, yields a good formula $H$ from $\vec{H}$ such that $\lfloor E \rfloor = \lfloor H \rfloor$ which, by (41), automatically means the **CL4**-provability of $\lfloor E \rfloor$. The remaining Subcase 1.3 is about when all surface occurrences of large (resp. medium) molecules in $E$ are negative (resp. positive). This will be shown to imply that $\lfloor E \rfloor$ follows from the floorifications of some elements of $\vec{H}$ by Rule **A** for "the same reasons as" $E$ follows from $\vec{H}$.

*Subcase 1.1:* $E$ has a positive surface occurrence of a large molecule $\check{P}_\sqcup^\sqcap(\vec{t})$. Pick any $a$ such that neither any $\check{P}_\sqcup^a$-molecule nor any $\check{P}_a^b$-molecule (whatever $b$) have independent occurrences in $E$. Such an $a$ exists, for otherwise we would have at least $m+1$ independent occurrences of molecules in $E$ (including the occurrence of $\check{P}_\sqcup^\sqcap(\vec{t})$), which contradicts **Cond1**. Let $H$ be the result of replacing in $E$ the above occurrence of $\check{P}_\sqcup^\sqcap(\vec{t})$ by $\check{P}_\sqcup^a(\vec{t})$. Clearly $H \in \vec{H}$. Observe that when transferring from $E$ to $H$, we just "downsize" $\check{P}_\sqcup^\sqcap(\vec{t})$ and otherwise do not create any additional independent occurrences of molecules, so **Cond1** continues to be satisfied for $H$. Neither do we introduce any new non-surface occurrences of molecules or any new independent occurrences of small molecules, so **Cond2** and **Cond3** also continue to hold for $H$. And our choice of $a$ obviously guarantees that so does **Cond4**. To summarize, $H$ is good. Therefore, by (41), $\mathbf{CL4} \vdash \lfloor H \rfloor$. Finally, note that, when floorifying a given formula, both $\check{P}_{\sqcup}^{\sqcap}(\vec{t})$ and $\check{P}_{\sqcup}^{a}(\vec{t})$ get replaced by the same atom $P(\vec{t})$; and, as the only difference between $E$ and $H$ is that $H$ has $\check{P}_{\sqcup}^{a}(\vec{t})$ where $E$ has $\check{P}_{\sqcup}^{\sqcap}(\vec{t})$, obviously $\lfloor H \rfloor = \lfloor E \rfloor$. Thus, $\mathbf{CL4} \vdash \lfloor E \rfloor$.

*Subcase 1.2:* $E$ has a negative surface occurrence of a medium molecule $\check{P}_{\sqcup}^{a}(\vec{t})$. Pick any $b$ such that $E$ does not have an independent occurrence of any $\check{P}_{b}^{a}$-molecule. Again, in view of **Cond1**, such a $b$ exists. Let $H$ be the result of replacing in $E$ the above occurrence of $\check{P}_{\sqcup}^{a}(\vec{t})$ by $\check{P}_{b}^{a}(\vec{t})$. Certainly $H \in \vec{H}$. Conditions **Cond1** and **Cond2** continue to hold for $H$ for the same reasons as in Subcase 1.1. In view of our choice of $b$, **Cond3** is also inherited by $H$ from $E$. And so is **Cond4**, because $H$ has the same positive occurrences of (the same) molecules as $E$ does. Thus, $H$ is good. Therefore, by (41), $\mathbf{CL4} \vdash \lfloor H \rfloor$. It remains to show that $\lfloor H \rfloor = \lfloor E \rfloor$. Note that when floorifying $E$, $\check{P}_{\sqcup}^{a}(\vec{t})$ gets replaced by $P(\vec{t})$. But so does $\check{P}_{b}^{a}(\vec{t})$ when floorifying $H$ because, by our choice of $b$, $\check{P}_{b}^{a}(\vec{t})$ is an isolated small molecule of $H$. Since the only difference between $H$ and $E$ is that $H$ has $\check{P}_{b}^{a}(\vec{t})$ where $E$ has $\check{P}_{\sqcup}^{a}(\vec{t})$, it is obvious that indeed $\lfloor H \rfloor = \lfloor E \rfloor$.

*Subcase 1.3:* None of the above two conditions is satisfied. This means that in $E$ all surface occurrences of large molecules are negative, and all surface occurrences of medium molecules are positive. Every large molecule $\check{P}_{\sqcup}^{\sqcap}(\vec{t})$ is a $\sqcap$-formula whose surface occurrences, as we remember, get replaced by $\top$ when transferring from $E$ to $\|E\|$; but the same happens to the corresponding occurrences of $P(\vec{t})$ in $\lfloor E \rfloor$ when transferring from $\lfloor E \rfloor$ to $\|\lfloor E \rfloor\|$ because, as we have just noted, such occurrences are negative, and negative surface occurences of general atoms get replaced by $\top$ when elementarizing **CL4**-formulas. Similarly, every medium molecule $\check{P}_{\sqcup}^{a}(\vec{t})$ is a $\sqcup$-formula so that its surface occurrences get replaced by $\bot$ when transferring from $E$ to $\|E\|$; but the same happens to the corresponding occurrences of $P(\vec{t})$ in $\lfloor E \rfloor$ when transferring from $\lfloor E \rfloor$ to $\|\lfloor E \rfloor\|$ because they are positive, and positive surface occurences of general atoms get replaced by $\bot$ when elementarizing **CL4**-formulas. Based on these observations, with a little thought we can see that $\|\lfloor E \rfloor\|$ is "almost the same" as $\|E\|$; specifically, the only difference between these two formulas is that $\|\lfloor E \rfloor\|$ has $\bot$ where $\|E\|$ has positive isolated (isolated in $E$ and hence in $\|E\|$) small molecules, and $\|\lfloor E \rfloor\|$ has $\top$ where $\|E\|$ has negative isolated small molecules. So, $\|\lfloor E \rfloor\|$ is the result of replacing in $\|E\|$ isolated molecules by $\bot$ or $\top$. Keeping this in mind, remember from classical logic — as we once already did in the proof of Lemma 3.2 — that replacing an isolated atom (an atom $p(\vec{t})$ such that letter $p$ has no other occurrences in the formula) by whatever formula preserves validity. Thus, if $\|E\|$ is classically valid, so is $\|\lfloor E \rfloor\|$. But $\|\lfloor E \rfloor\|$ indeed is classically valid because $E$ is derived by Rule **A**. We conclude that

$$\lfloor E \rfloor \text{ is stable.} \tag{42}$$

**Subclaim 1.1**

(i) *Whenever $\lfloor E \rfloor$ has a positive (resp. negative) quasiatom $K$ of the form $G_1 \sqcap \ldots \sqcap G_n$ (resp. $G_1 \sqcup \ldots \sqcup G_n$), for each $i \in \{1, \ldots, n\}$, there is a good formula $H$ in $\vec{H}$ such that $\lfloor H \rfloor$ is the result of replacing $K$ by $G_i$ in $\lfloor E \rfloor$.*

(ii) *Whenever $\lfloor E \rfloor$ has a positive (resp. negative) quasiatom $K$ of the form $\sqcap x G(x)$ (resp. $\sqcup x G(x)$), there is a good formula $H$ in $\vec{H}$ such that $\lfloor H \rfloor$ is the result of replacing $K$ by $G(y)$ in $\lfloor E \rfloor$, where $y$ is a variable not occurring in $\lfloor E \rfloor$.*

For clause (i) of the subclaim, assume $\lfloor E \rfloor$ has a positive (resp. negative) quasiatom $K = G_1 \sqcap \ldots \sqcap G_n$ (resp. $K = G_1 \sqcup \ldots \sqcup G_n$), and consider any $i \in \{1, \ldots, n\}$. The logical structure of $E$ is the same as that of $\lfloor E \rfloor$, with the only difference that, wherever $\lfloor E \rfloor$ has general atoms, $E$ has molecules. Hence, where $\lfloor E \rfloor$ has the above $K$, $E$ has a positive (resp. negative) occurrence of a quasiatom $K' = G_1' \sqcap \ldots \sqcap G_n'$ (resp. $K' = G_1' \sqcup \ldots \sqcup G_n'$). Then, since $E$ is derived from $\vec{H}$ by Rule **A**, $\vec{H}$ contains the result $H$ (fix it) of replacing $K'$ by $G_i'$ in $E$. It is not hard to see that $\lfloor H \rfloor$ is the result of replacing $K$ by $G_i$ in $\lfloor E \rfloor$. What remains to show now is that $H$ is good. When transferring from $E$ to $H$, **Cond1** is inherited by $H$ for the same reason as in all of the previous cases. So is **Cond2** because we are not creating any new non-surface occurrences. Furthermore, since $K'$ is not a molecule (for otherwise $\lfloor E \rfloor$ would have a general atom instead of $K$), **Cond2** guarantees that $K_i'$ is not a small or medium molecule. This means that, when transferring from $E$ to $H$, we are not creating new (nor destroying old) independent/surface occurrences of any small or medium molecules, so that **Cond3** and **Cond4** are also inherited by $H$ from $E$. To summarize, $H$ is indeed good, and clause (i) of Subclaim 1.1 is thus proven. Clause (ii) can be handled in a similar way.

Now (42), (41) and Subclaim 1.1 immediately imply that $\lfloor E \rfloor$ is derivable in **CL4** by Rule **A**. We are done with Subcase 1.3 and hence CASE 1.

The remaining two CASES 2 and 3 are about when $E$ is derived by one of the Rules **B1** or **B2** from a premise $H$. Such a $H$ turns out to be good and hence (by the induction hypothesis) its floorification **CL4**-provable. And, "almost always" $\lfloor E \rfloor$ follows from $\lfloor H \rfloor$ by **B1** or **B2** for the same reasons as $E$ follows from $H$. An exception is when $H$ is the result of replacing in $E$ a positive occurrence of a medium molecule $\check{P}^a_{\sqcup}(\vec{t})$ by one of its disjuncts $\check{P}^a_b(\vec{t})$ (so that we are talking about an application of Rule **B1**) such that $E$ has a negative independent occurrence of a $\check{P}^a_b$-molecule. Using our earlier terms, this is a step signifying $\top$'s (final) decision to "match" the two $P$-based molecules. In this case, while $\lfloor E \rfloor$ does not follow from $\lfloor H \rfloor$ by Rule **B1**, it does so by Rule **C**. The secret is that the two $P$-based molecules are non-isolated small molecules in $H$ and hence remain elementary atoms in $\lfloor H \rfloor$, while they turn into general atoms in $\lfloor E \rfloor$.

*CASE 2:* $E$ is derived by Rule **B1**. That is, we have **CL3** $\vdash H$, where $H$ is the result of replacing in $E$ a negative (resp. positive) quasiatom $K$ of the form $G_1 \sqcap \ldots \sqcap G_n$ (resp. $G_1 \sqcup \ldots \sqcup G_n$) by $G_i$ for some $i \in \{1, \ldots, n\}$. Fix these formulas and this number $i$. Just as in CASE 1 (statement (41)), based on the induction hypothesis, we have:

$$\text{If } H \text{ is good, then } \textbf{CL4} \vdash \lfloor H \rfloor. \tag{43}$$

We need to consider the following three subcases that cover all possibilities:

*Subcase 2.1:* $K$ is not a molecule. Reasoning (almost) exactly as we did in the proof of Subclaim 1.1(i), we find that $H$ is good. Therefore, by (43), **CL4** $\vdash \lfloor H \rfloor$. Now, a little thought can convince us that $\lfloor E \rfloor$ follows from $\lfloor H \rfloor$ by Rule **B1**, so that **CL4** $\vdash \lfloor E \rfloor$.

*Subcase 2.2:* $K$ is a large molecule $\check{P}^{\sqcap}_{\sqcup}(\vec{t})$. So, $K$ is negative in $E$, and $G_i = \check{P}^i_{\sqcup}(\vec{t})$. A (now already routine for us) examination of **Cond1**-**Cond4** reveals that each of these four conditions are inherited by $H$ from $E$, so that $H$ is good. Therefore, by (43), **CL4** $\vdash \lfloor H \rfloor$. Now, $\lfloor H \rfloor$ can be easily seen to be the same as $\lfloor E \rfloor$, and thus **CL4** $\vdash \lfloor E \rfloor$.

*Subcase 2.3:* $K$ is a medium molecule $\check{P}^a_{\sqcup}(\vec{t})$. So, $K$ is positive in $E$, and $G_i = \check{P}^a_i(\vec{t})$. There are two subsubcases to consider:

*Subsubcase 2.3.1:* $E$ contains no independent occurrence of a $\check{P}^a_i$-molecule. One can easily verify that $H$ is good and that $\lfloor H \rfloor = \lfloor E \rfloor$. By (43), we then get the desired **CL4** $\vdash \lfloor E \rfloor$.

*Subsubcase 2.3.2:* $E$ has an independent occurrence of a $\check{P}^a_i$-molecule $\check{P}^a_i(\vec{t'})$. Since $E$ also has a positive independent occurrence of $\check{P}^a_{\sqcup}(\vec{t})$, **Cond4** implies that the above occurrence of $\check{P}^a_i(\vec{t'})$ in $E$ is negative. This, in conjunction with **Cond3**, means that $E$ does not have any other independent occurrences of $\check{P}^a_i$-molecules, and thus $H$ has exactly two — one negative and one positive — independent occurrences of $\check{P}^a_i$-molecules. This guarantees that **Cond3** is satisfied for $H$, because $H$ and $E$ only differ in that $H$ has $\check{P}^a_i(\vec{t})$ where $E$ has $\check{P}^a_{\sqcup}(\vec{t})$. Conditions **Cond1** and **Cond2** are straightforwardly inherited by $H$ from $E$. Finally, **Cond4** also transfers from $E$ to $H$ because, even though $H$ — unlike $E$ — has a positive independent occurrence of a $\check{P}^a_i$-molecule, it no longer has a positive independent occurrence of a $\check{P}^a_{\sqcup}$-molecule (which, by the same condition **Cond4** for $E$, was unique in $E$). Thus, $H$ is good and, by (43), **CL4** $\vdash \lfloor H \rfloor$. Note that since $H$ is good, by **Cond2**, the independent occurrences of $\check{P}^a_i(\vec{t'})$ and and $\check{P}^a_i(\vec{t})$ in it are surface occurrences. The same, of course, is true for the corresponding occurrences of $\check{P}^a_i(\vec{t'})$ and $\check{P}^a_{\sqcup}(\vec{t})$ in $E$. Let us now compare $\lfloor E \rfloor$ with $\lfloor H \rfloor$. According to our earlier observation, there is only one independent occurrence of a $\check{P}^a_i$-molecule — specifically, $\check{P}^a_i(\vec{t'})$ — in $E$, i.e. $\check{P}^a_i(\vec{t'})$ is $E$-isolated. Hence, when floorifying $E$, the independent occurrence of $\check{P}^a_i(\vec{t'})$ gets replaced by $P(\vec{t'})$ and the independent occurrence of $\check{P}^a_{\sqcup}(\vec{t})$ replaced by $P(\vec{t})$. On the other hand, $\check{P}^a_i(\vec{t'})$ is no longer isolated in $H$, so $\check{P}^a_i(\vec{t'})$ stays as it is when floorifying $H$; so does $\check{P}^a_i(\vec{t})$ because it is not isolated in $H$, either. Based on this observation, we can easily see that the only difference between $\lfloor E \rfloor$ and $\lfloor H \rfloor$ is that $\lfloor E \rfloor$ has $P(\vec{t})$ and $P(\vec{t'})$ where $\lfloor H \rfloor$ has $\check{P}^a_i(\vec{t})$ and $\check{P}^a_i(\vec{t'})$. Since $\lfloor E \rfloor$ does not contain $\check{P}^a_i$ (because the only independent occurrence of a $\check{P}^a_i$-molecule got replaced by a general atom when floorifying $E$), and since we deal with two — one positive and one negative — surface occurrences of $P$-based general atoms in $\lfloor E \rfloor$, we find that $\lfloor E \rfloor$ follows from $\lfloor H \rfloor$ by Rule **C**. We already know that **CL4** $\vdash \lfloor H \rfloor$. Hence

$\mathbf{CL4} \vdash \lfloor E \rfloor$.

*CASE 3:* $E$ is derived by Rule **B2**. That is, we have $\mathbf{CL3} \vdash H$, where $H$ is the result of replacing in $E$ a negative (resp. positive) quasiatom $K$ of the form $\sqcap xG(x)$ (resp. $\sqcup xG(x)$) by $G(t)$, with the known restrictions on term $t$. This case is similar to CASE 2, but is, in fact, much simpler as $K$ cannot be a molecule, so we only deal with a situation corresponding to Subcase 2.1 of the previous case. The induction hypothesis, as before, implies statement (43). A routine examination of **Cond1-Cond4** reveals that these conditions are inherited by $H$ from $E$, so that, by (43), $\mathbf{CL4} \vdash \lfloor H \rfloor$. Now, $\lfloor E \rfloor$ can be easily seen to follow from $\lfloor H \rfloor$ by Rule **B2**, so that $\mathbf{CL4} \vdash \lfloor E \rfloor$.

Claim 1 is proven.

Now we are close to finishing our proof of Lemma 5.1. Assume $\mathbf{CL4} \not\vdash F$. Let $\lceil F \rceil$ be the result of replacing in $F$ all occurrences of each ($n$-ary) general atom $P(t_1, \ldots, t_n)$ by $\check{P}_\sqcup^\sqcap(t_1, \ldots, t_n)$. Obviously $\lceil F \rceil$ is good. Clearly we also have $\lfloor \lceil F \rceil \rfloor = F$, so that $\mathbf{CL4} \not\vdash \lfloor \lceil F \rceil \rfloor$. Therefore, by Claim 1, $\mathbf{CL3} \not\vdash \lceil F \rceil$. Hence, by Theorem 5.9(b) of [6], there is an $\lceil F \rceil$-admissible interpretation $\dagger$ that interprets every elementary letter as a finitary predicate of arithmetical complexity $\Delta_2$, such that

$$\not\models \lceil F \rceil^\dagger. \tag{44}$$

Since $\dagger$ interprets all letters of $\lceil F \rceil$ as finitary predicates, we may assume that none of those predicates depend on any of the variables from our earlier-fixed pool $z_1, z_2, \ldots$ (otherwise, take any other variables instead). Let $*$ be the interpretation[17] that agrees with $\dagger$ on all elementary letters, and interprets each $n$-ary general letter $P \in \mathcal{P}$ as $P^*(z_1, \ldots, z_n)$ defined by

$$P^*(z_1, \ldots, z_n) = \left(\check{P}_\sqcup^\sqcap(z_1, \ldots, z_n)\right)^\dagger.$$

Thus, $*$ interprets letters as promised in Lemma 5.1. To complete our proof, it remains to show is that $*$ is $F$-admissible and $\not\models F^*$. In showing admissibility, we only need to care about general letters, because elementary letters are taken care of by the fact that on them $*$ agrees with the $\lceil F \rceil$-admissible $\dagger$.

Consider any general atom $P(t_1, \ldots, t_n) = P(\vec{t})$ of $F$. Based on the $\lceil F \rceil$-admissibility of $\dagger$ and our assumption that the variables $\vec{z} = z_1, \ldots, z_n$ are sufficiently "neutral", the following two statements can be verified by a routine analysis of the relevant definitions:

$$P^*(\vec{z}) \text{ does not depend on any variables that are not among } \vec{z} \text{ but occur in } F; \tag{45}$$
$$P^*(\vec{t}) = \left(\check{P}_\sqcup^\sqcap(\vec{t})\right)^\dagger. \tag{46}$$

By (45), condition (i) of Definition 2.1 is satisfied. Next, as pointed out in [6], for every **CL3**-formula $E$ and interpretation $\star$, the game $E^\star$ is unistructural. So $P^*$, which is $\left(\check{P}_\sqcup^\sqcap(z_1, \ldots, z_n)\right)^\dagger$, is unistructural and hence, of course, unistructural in all variables. This means that condition (ii) of Definition 2.1 is also satisfied, and we conclude that $*$ is $F$-admissible.

Finally, based on (46), we immediately find that $F^* = \lceil F \rceil^\dagger$. This, by (44), implies $\not\models F^*$. $\square$

## 6 Decidability of the $\forall, \exists$-free fragment of CL4

This section is devoted to a proof of Theorem 2.5. Here we let $F$ range over **CL4**-formulas not containing blind quantifiers. The decidability of the question $\mathbf{CL4} \vdash F$ can be shown by induction on the **aggregate complexity** of $F$, by which we mean the number of occurrences of logical operators plus the number of occurrences of general atoms in $F$.

$F$ is provable iff it is derivable from some provable formulas by one of the Rules **A**, **B1**, **B2** or **C**. Our decision procedure is a recursive one that tests, in turn, each of these four possibilities. If one of those four tests succeeds, the procedure returns "yes", otherwise returns "no".

---

[17]For full accuracy, we should have said "an interpretation" here, because — out of laziness — we are not defining $*$ for general letters not occurring in $F$. Earlier in this section we agreed to pretend that those letters simply do not exist. Such letters are indeed irrelevant and, if there was any need, our $*$ could be extended to them in an arbitrary way.

**Testing Rules A, B1 and B2:** These three tests are done in the same way as described in Section 10 of [6] for the $\forall, \exists$-free fragment of **CL3**, and there is no need to (almost) literally reproduce that description here. The fact that we deal with the more expressive language of **CL4** hardly creates any differences: the question of stability of a $\forall, \exists$-free formula still remains decidable, and Lemma 9.1 of [6] on which some of our old arguments relied is true for **CL4**-formulas for virtually the same reasons as it was true for **CL3**-formulas.

**Testing Rule C:** For each pair consisting of one positive and one negative surface occurrence of some $n$-ary general letter $P$ in $F$, do the following: pick an arbitrary (say, the lexicographically smallest) $n$-ary non-logical elementary letter $q$ not occurring in $F$, replace in $F$ the above two occurrences of $P$ by $q$, and see if the resulting formula $H$ is **CL4**-provable. The aggregate complexity of $H$ is lower than that of $F$, so, by the induction hypothesis, testing $H$ for **CL4**-provability can be done effectively. Also, there is only a finite number of such $H$s to test. This is so because $F$ only has finitely many (pairs of occurrences of) general atoms $P$, and the above selection of a particular $q$ is clearly irrelevant: the outcome would be the same if any other — not occurring in $F$ — $n$-ary non-logical elementary letter $q'$ was selected, for there would be virtually no reason why a formula would be provable with $q$ but not with $q'$. So, the whole procedure takes a finite amount of time. If one of the above $H$s turns out to be **CL4**-provable, the step is considered to have succeeded. Otherwise it has failed. Of course our test is correct: it succeeds if and only if $F$ is derivable by Rule **C**.

As an aside, an analysis the above construction can easily convince us that:

**Fact 6.1** *Our decision algorithm for the $\forall, \exists$-free fragment of **CL4** runs in polynomial space.*

## 7 Appendix A: Proof of Fact 2.2

Consider a **CL4**-formula $F$ and an $F$-admissible interpretation $*$. We want to see that the game $F^*$ is defined, which can be done by induction on the complexity of $F$. The basis case of induction is trivial. And the only nontrivial case in the inductive step is when the main operator of $F$ is a blind quantifier. This is so because all operations except blind quantification are defined for all problems. So, assume $F$ is $\forall y G$ or $\exists y G$ where, by the induction hypothesis, the game $G^*$ is defined. Then $F^*$ is defined iff $G^*$ is unistructural in $y$. In order to show that $G^*$ is $y$-unistructural, it would suffice to verify that, for every atom $E$ of $G$, $E^*$ is $y$-unistructural. This is so because, according to Theorem 14.1 of [2], each of the operations $\neg, \wedge, \vee, \sqcap, \sqcup, \forall, \exists, \sqcap, \sqcup$ preserves the $y$-unistructural property.

Pick any atom $E$ of $G$. If $E$ is an elementary atom, then $E^*$, as an elementary game, is simply unistructural and hence $y$-unistructural. Now, for the rest of this proof, suppose $E$ is an ($n$-ary) general atom $P(t_1, \ldots, t_n)$, with $P$ interpreted as $P^*(x_1, \ldots, x_n)$. Our goal is to show that $P^*(t_1, \ldots, t_n)$ is $y$-unistructural.

Without loss of generality, we may assume that, for some $i$ (fix it) with $0 \leq i \leq n$, we have $t_1 = \ldots = t_i = y$ and $y \notin \{t_{i+1}, \ldots, t_n\}$. Then, by clause (ii) of Definition 2.1,

$$P^*(x_1, \ldots, x_n) \text{ is unistructural in each of the variables } x_1, \ldots, x_i. \tag{47}$$

Consider any two valuations $e_1$ and $e_2$ that agree on all variables but $y$. We want to verify that

$$\mathbf{Lr}^{e_1[P^*(t_1,\ldots,t_n)]} = \mathbf{Lr}^{e_2[P^*(t_1,\ldots,t_n)]}. \tag{48}$$

Let $e_1'$ (resp. $e_2'$) be the valuation that agrees with $e_1$ (resp. $e_2$) on all variables that are not among $x_1, \ldots, x_n$, and sends each $x_j$ ($1 \leq j \leq n$) to $e_1(t_j)$ (resp. $e_2(t_j)$). By Definition 4.1 of [6],

$$e_1[P^*(t_1,\ldots,t_n)] = e_1'[P^*(x_1,\ldots,x_n)] \text{ and } e_2[P^*(t_1,\ldots,t_n)] = e_2'[P^*(x_1,\ldots,x_n)]. \tag{49}$$

**Claim 1** *All variables on which $e_1'$ and $e_2'$ may disagree are among $\{x_1, \ldots, x_i, y\}$; besides, if $e_1'$ and $e_2'$ (really) disagree on $y$, then $P^*(x_1, \ldots, x_n)$ is unistructural in $y$.*

To verify this claim, consider any variable $z$ on which $e_1'$ and $e_2'$ disagree. By our assumption, each $t_j$ with $i < j \leq n$ is different from $y$, so $e_1(t_j) = e_2(t_j)$ and hence $e_1'(x_j) = e_2'(x_j)$. This means that $z$ — on which $e_1'$ and $e_2'$ disagree — cannot be in $\{x_{i+1}, \ldots, x_n\}$. So, if $z \in \{x_1, \ldots, x_n\}$, then $z$ must be in $\{x_1, \ldots, x_i\}$.

And, if here $z = y$, then, by (47), $P^*(x_1, \ldots, x_n)$ is unistructural in $y$. Now suppose $z \notin \{x_1, \ldots, x_n\}$. Unless $z = y$, $e_1$ and $e_2$ agree on $z$ and hence so do $e_1'$ and $e_2'$, which is a contradiction. So, $z = y$. Also, since $y \notin \{x_1, \ldots, x_n\}$ and $y$ occurs in $F$, condition (i) of Definition 2.1 implies that the game $P^*(x_1, \ldots, x_n)$ does not depend on $y$ and hence is $y$-unistructural. Claim 1 is proven.

Combining (47) with Claim 1, we find that $e_1'$ and $e_2'$ only disagree on at most finitely many variables, and the game $P^*(x_1, \ldots, x_n)$ is unistructural in every variable on which $e_1'$ and $e_2'$ disagree. With a little thought we can see that then $\mathbf{Lr}^{e_1'[P^*(x_1,\ldots,x_n)]} = \mathbf{Lr}^{e_2'[P^*(x_1,\ldots,x_n)]}$ which, by (49), implies $\mathbf{Lr}^{e_1[P^*(t_1,\ldots,t_n)]} = \mathbf{Lr}^{e_2[P^*(t_1,\ldots,t_n)]}$. This completes our proof of (48) and hence of Fact 2.2.

# 8 Appendix B: Proof of Fact 3.8

Assume $F$ is any hyperformula, $*$ is an $F$-admissible interpretation, and $\star$ is the perfect interpretation induced by $(*, e)$. We want to show that then $\star$ is also $F$-admissible. Just as in the proof of Lemma 3.9, we may also assume that $F$ does not contain hybrid atoms, for otherwise replace $F$ with its general dehybridization. Since $\star$ is a perfect interpretation, it automatically satisfies clause (i) of Definition 2.1. To see that clause (ii) is also satisfied, pick any $n$-ary letter $L$ of $F$, interpreted as $L^*(x_1, \ldots, x_n)$ by $*$ and as $L^\star(x_1, \ldots, x_n)$ by $\star$. It would suffice to show that whenever $L^*$ is unistructural in $x_i$ ($1 \leq i \leq n$), so is $L^\star$. For simplicity, let us just consider the case $i = 1$. Assume $L^\star$ is not unistructural in $x_1$, i.e. there are two valuations $f$ and $g$ that agree on all variables but $x_1$ such that $\mathbf{Lr}^{f[L^\star]} \neq \mathbf{Lr}^{g[L^\star]}$. Let $a = f(x_1)$, $b = g(x_1)$, $c_2 = f(x_2) = g(x_2)$, ..., $c_n = f(x_n) = g(x_n)$. Then clearly $f[L^\star] = L^\star(a, c_2, \ldots, c_n)$ and $g[L^\star] = L^\star(b, c_2, \ldots, c_n)$, so that $\mathbf{Lr}^{L^\star(a,c_2,\ldots,c_n)} \neq \mathbf{Lr}^{L^\star(b,c_2,\ldots,c_n)}$. From here, taking into account that $\star$ is the perfect interpretation induced by $(*, e)$ and hence $L^\star(a, c_2, \ldots, c_n) = e[L^*(a, c_2, \ldots, c_n)]$ and $L^\star(b, c_2, \ldots, c_n) = e[L^*(b, c_2, \ldots, c_n)]$, we find $\mathbf{Lr}^{e[L^*(a,c_2,\ldots,c_n)]} \neq \mathbf{Lr}^{e[L^*(b,c_2,\ldots,c_n)]}$. The latter can be obviously rewritten as $\mathbf{Lr}^{L^*}_{e_1} \neq \mathbf{Lr}^{L^*}_{e_2}$, where $e_1$ and $e_2$ are the valuations with $e_1(x_1) = a$, $e_2(x_1) = b$, $e_1(x_2) = e_2(x_2) = c_2$, ..., $e_1(x_n) = e_2(x_n) = c_n$, and $e_1(y) = e_2(y) = e(y)$ for every variable $y \notin \{x_1, \ldots, x_n\}$. Since $e_1$ and $e_2$ agree on all variables but $x_1$, we find that $L^*$ is not $x_1$-unistructural.

# Index

**Nonalphabetical symbols and notation:**